\documentclass[12pt]{article}
\usepackage{amsmath}
\usepackage{epsfig}
\usepackage{here}
\usepackage{afterpage}
\begin{document}
\begin{titlepage}
\begin{center}
\vspace*{1.0cm}
\begin{tabbing}
xxxxxxxxxxxxxxxxxxxxxxxxxxxxxxxxxxxxxxxxxxxxxxxxxxxxxxxxxxxx \= \kill
\> {\bf NuMI-B-781} \\
\end{tabbing}
\vskip 2cm

{\LARGE \bf Neutrino Spectrum at the Far Detector\\
\vspace*{6mm}
 Systematic Errors}\\
\vskip 1cm
    
\vspace*{5mm}
{\it \large \today}\\ 
Version 2.2 \\
\vspace*{1.0cm}
M. Szleper, Northwestern University

\vspace*{5mm}
A. Para, Fermilab

\end{center}
\vskip 2cm
\begin{abstract}
Neutrino oscillation experiments often employ two identical detectors to minimize
errors due to inadequately known neutrino beam. We examine various systematics
effects related to the prediction of the neutrino spectrum in the `far' detector
on the basis of the spectrum observed at the `near' detector. We propose a novel 
method of the derivation of the far detector spectrum. This method is less sensitive
to the details of the understanding of the neutrino beam line and the hadron
production spectra than the usually used `double ratio' method thus allowing to
reduce the systematic errors. 
\end{abstract}
\end{titlepage}
\newpage
\newpage
\tableofcontents
\newpage
\listoffigures

\newpage

\section{Introduction}

Neutrino oscillations experiments can be divided into two classes:
\begin{itemize}
\item disappearance experiments

In these experiments one  measures a deficit of the observed rate of neutrino
interactions with respect to the rate expected from the knowledge of the
neutrino beam.

\item appearance experiments

In these experiments one detects interactions of the neutrino flavor not present
in the neutrino beam at its source

\end{itemize}

A positive result of an appearance experiment constitutes an undisputable
proof of the existence of the neutrino oscillations and it is fairly 
independent of the detailed understanding of the neutrino beam. This is often
contrasted with the disappearance experiments, where the uncertainty
of the neutrino flux at the position of the detector usually dominates
the systematic error. 

Further steps of the determination of the oscillation parameters involve a 
quantitative estimate of the rate of appearance or disappearance and their 
dependence on the neutrino energy. At this stage the understanding of the
neutrino flux at the position of the detector constitutes a major source
of the systematic error.

Absolute prediction of the neutrino fluxes is difficult, as the saga of 
measuring of the total cross sections teaches us. Modern tools of detailed
simulations make the predictions somewhat more reliable, but accuracy of
these predictions is limited by the scarcity of the data constraining the
shower simulation code. 

A powerful method of improving the knowledge of  the neutrino flux at the
position of the detector consists using two detectors in the same beam line:
\begin{itemize}
\item a `near' detector positioned near the neutrino source to establish
the beam characteristics and to enable a reliable prediction of the neutrino
flux at the 'far' detector

\item a `far' detector positioned at the distance where oscillation effects
are expected to be present.
\end{itemize}

This technique was pioneered by the CCFR\cite{CCFR} collaboration at Fermilab 
and then subsequently used by CDHSW\cite{CDHSW} and CHARM\cite{CHARM} experiments
at CERN. Recently the two detector approach was employed by the K2K\cite{K2K}
experiment in Japan.

High statistics of the neutrino interactions expected at MINOS offer
a potential for precise determination of the oscillations parameters. This,
in turn, requires  a very good understanding of the neutrino flux.

This note describes contributions to the systematic errors of the
neutrino flux prediction and proposes a novel method of predicting the neutrino
flux with reduced sensitivity to various systematic effects.

\section{Neutrino Beam}

Neutrino beam is produced in three steps:
\begin{itemize}
\item pions and kaons are produced in the target
\item (optional, but necessary to attain high beam intensities) produced 
particles are focused directed towards the detector
\item pions and kaons decay inside the decay volume producing neutrinos
\end{itemize}

We examine these steps at some level of details to indicate the possible 
sources of systematic errors of the neutrino flux predictions.
 
\subsection{Particle production in the target}

Primary protons strike the target and produce pions and kaons. For low
energy beams and long baseline experiments (hence neutrinos produced
at small angles) pions produce the main  contribution to the neutrino
flux.

Production spectra of pions produced by protons are known rather
poorly. Surprisingly large uncertainties exist in the knowledge
of the inclusive cross sections, $E \frac{d^3\sigma}{dp^3}$, even
for the elementary $pp$ reaction. Situation is further complicated by
the fact that the neutrino beams utilize extended targets, where
the absorption and re-interaction of the particles produced in the
first interaction play a significant role.

Uncertainty in the production spectra is a direct consequence of
a scarcity of the relevant experimental data. The primary source
of the experimental information is the measurement of the particles
yields in $pBe$ collisions at 400 and 450 GeV/c 
performed at CERN as a tool for understanding
the CERN NBB and WBB neutrino beams\cite{Atherton,SPY1,SPY2}. Extrapolation
to different target materials is guided by the measurement of Barton
et al.\cite{Barton} 

There are two approaches to predictions of the particle fluxes for the
neutrino beam:
\begin{itemize}
\item use the existing data to constrain and tune complete shower 
simulation codes. 

This approach is represented by FLUKA \cite{FLUKA}
and MARS \cite{MARS}. It should be pointed out that we use 
FLUKA as implemented in GEANT (often referred as GFLUKA). There
exists a more modern version of FLUKA, re-tuned using the SPY\cite{SPY1,SPY2}
data, but unfortunately it is not interfaced to GEANT yet.

\item parameterize the existing data with an ad-hoc analytical formula
and use it as a particles source function. Some physics-driven scaling
laws must be employed to enable extrapolation of the data to different
targets, proton energies and phase space regions of the produced particles.

This approach is represented by Malensek\cite{Malensek} and BMPT\cite{BMPT}
`models'. The BMPT parameterization covers a complete phase space of the
produced particles. It is  based on much more complete data set and it is
developed for predictions of neutrino beam produced by different energy protons. 
Malensek formula is an {\it interpolation} of the Atherton\cite{Atherton}
data derived for a specific case of the Fermilab NBB. Use of this formula
outside the region covered by the data points used is ill justified.
We use it here, anyway, as an example of rather extreme variation of the production 
spectra.

\end{itemize}

The primary difference between various production models is the absolute
flux predictions. A typical range of the variation is of the order of $20 \%$.
Whereas this is an important practical factor, it is not important
for the prediction of the shape of the spectrum observed at the far
detector. 

A potential  source of the systematic error on the shape of the far detector
spectrum is related to the shape of the $p_{t}$ distribution of the
produced particles. The shape of the $p_{t}$ determines the angular divergence
of the hadron beam after the focusing elements and consequently the
distribution of the decay points of pions along the decay volume.

\begin{figure}[h]
\centerline{\epsfig{file=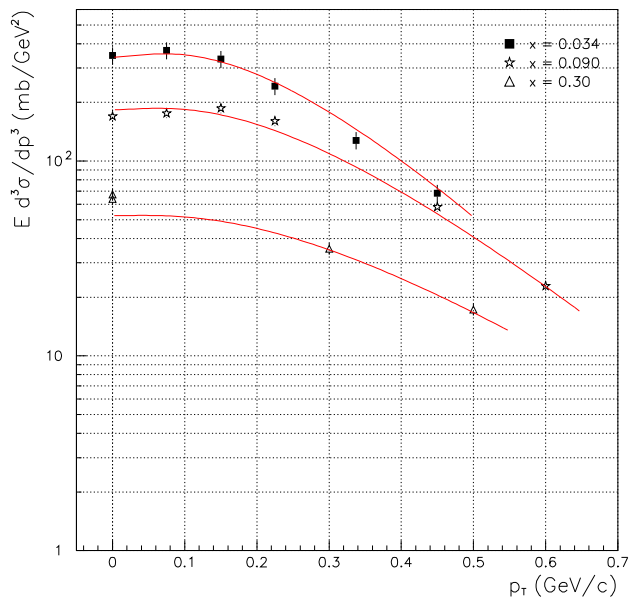,width=9cm}}
\caption{Invariant cross section for $\pi^{+}$ production as 
a function of $p_T$. Lines represent prediction of the
BMPT model for different longitudinal momenta of $\pi^{+}$.
Data points are from the Refs.\cite{Atherton,SPY1,SPY2}} 
\label{comp_BMPT}
\end{figure}

\begin{figure}[h]
\centerline{\epsfig{file=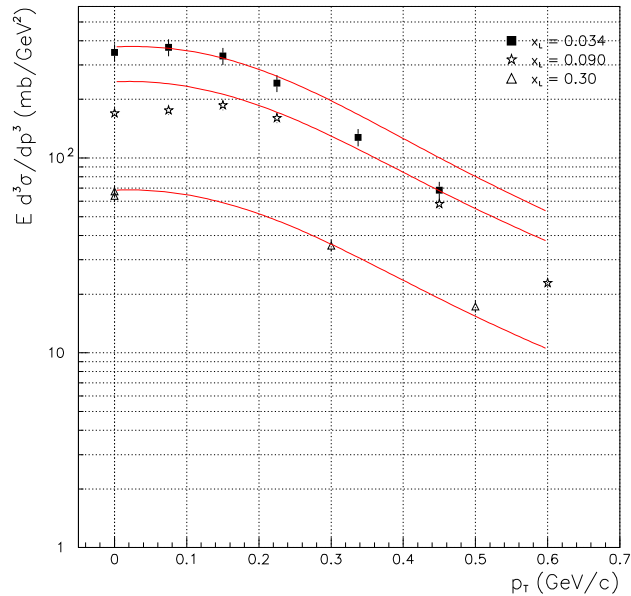,width=9cm}}
\caption{Invariant cross section for $\pi^{+}$ production as 
a function of $p_T$. Lines represent prediction of the
MARS model for different longitudinal momenta of $\pi^{+}$.
Data points are from the Refs.\cite{Atherton,SPY1,SPY2} }
\label{comp_MARS}
\end{figure}

Figs.\ref{comp_BMPT} and \ref{comp_MARS} compare the shape of the
$p_{t}$ spectra predicted by the BMPT and MARS models \cite{Alberto}.
They show that a considerable uncertainty exists in the shape:
physics-inspired general models tend to have an exponential fall-off
with $p_{t}$, whereas the experimental parameterization follows the
experimental data better.

A relationship between the $p_{t}$ distribution of produced pions
and the neutrino spectra is illustrated in Fig.\ref{enu_pt} by showing
the distributions of $p_{t}$ of pions responsible for neutrinos
of different energies, as predicted by different production models.

\afterpage{\clearpage
\begin{figure}[h]
\centerline{\epsfig{file=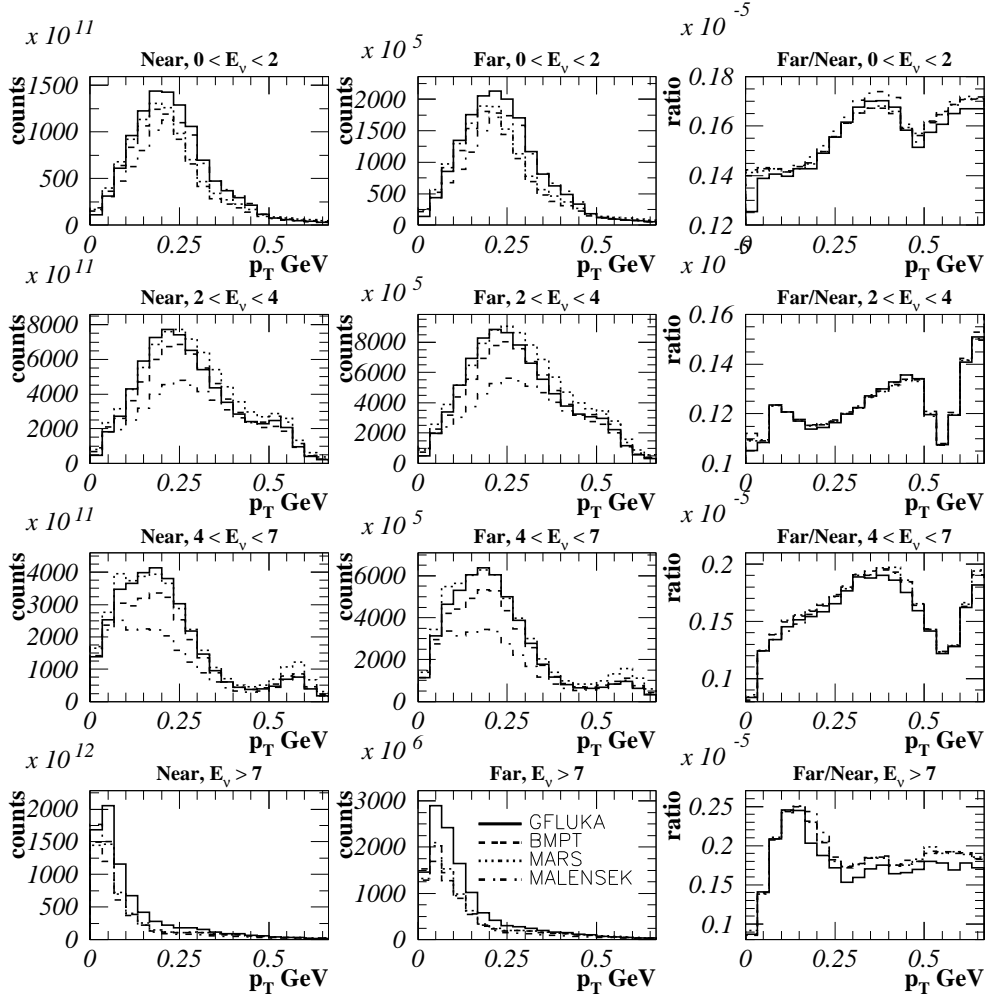,width=13cm}}
\caption{Distributions of the production $p_t$ of neutrino parents
weighted by the neutrino flux detected in the near (left column) 
and far (center column) detectors, predicted by 
different hadron production models.
Ratio of the $p_{t}$ distributions producing neutrinos of different energies
in the far and near detectors is shown in the right column.}
\label{enu_pt}
\end{figure}
}

Systematic errors due to hadron production rates on the predicted
neutrino fluxes have been extensively studied recently in general
context\cite{BMPT}, as well as in the context of NuMI/MINOS experiment
\cite{hadron,myhadr}.

\subsection{Focusing Elements}

A system of parabolic magnetic horns focuses the secondary particles by
giving them a $p_{t}$ kick proportional  
to the radius at which particle traverses the 
horn.
In case of a pointlike source  such a system focuses all 
particles produced with a particular momentum.
Finite radial sizes of the
 horns lead to a finite momentum byte of the secondary particles
which are focused. The central momentum of the focused particles
depends on the relative distance of the target and the horn.  

The hadron beam does have a significant angular divergence due to several
effects:
\begin{itemize}

\item particles produced at small angles  pass through
the opening of the magnetic horns and enter the decay volume with the
angular divergence characteristic for the bare target beam. This component is
defined by the aperture of the horn opening and the distance of the target
from the horn. These particles produce a high energy tail of the neutrino beam. 
This component
produces also the main contribution  of the  $\overline {\nu_{\mu}}$ 
background.

\item  finite size of the primary proton beam 

\item the finite size
of the production target.  Particles produced upstream of the focal point
the the first horn  will be overfocused whereas particles produced downstream of the focus will be underfocused, thus producing a 
divergent beam.  
\end{itemize}

\begin{figure}[h]
\centerline{\epsfig{file=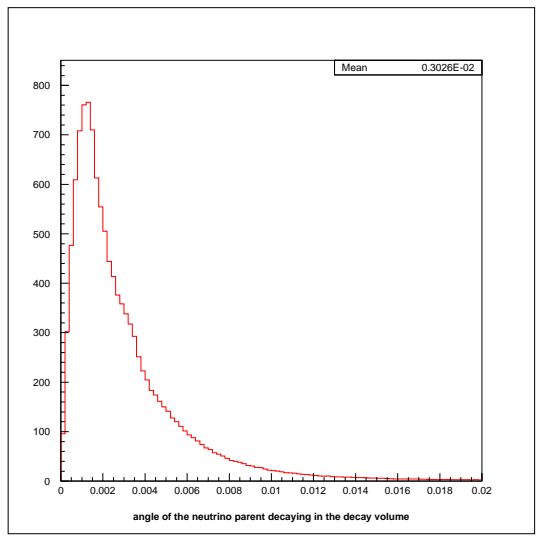,height=7cm}} 
\caption{Angular distribution of neutrino parents at the entrance to the
decay pipe.}
\label{angular_all}
\end{figure}

An overall angular distribution of hadrons after the focusing system 
is shown in the  Fig.\ref{angular_all}.  

\subsection{Finite target length}
 
The target length is typically comparable to the
distance between the target and the first horn. As the result, different
parts of the parent momentum spectrum will fall into acceptance of the 
horn for different  sections of the target: higher momentum particles from the 
the beginning of the target and lower momentum particles from the end of 
the target will be focused. This will result in the difference in the 
spectra of the resulting neutrinos, as shown in the Fig.\ref{targ_spect}.
The three effects are readily apparent:
\begin{itemize}

\item A number of pions (and hence a number of neutrinos) produced decreases 
along the target. This is due to the attenuation of the primary proton along 
the target.  

\item Spectrum of pions (and neutrinos) is shifted towards lower energies
along the target. This is the effect of the finite acceptance of the horn 
focusing system.

\item Relative contribution of the high energy tail increases along the target.
This tail is due to the small angle particles which pass through the opening
in the horn. The angular acceptance is much smaller for particles produced 
in the front section of the target. The attenuation of small angle particles 
traversing the target contributes another suppression factor.
  
\begin{figure}[h]
\centerline{\epsfig{file=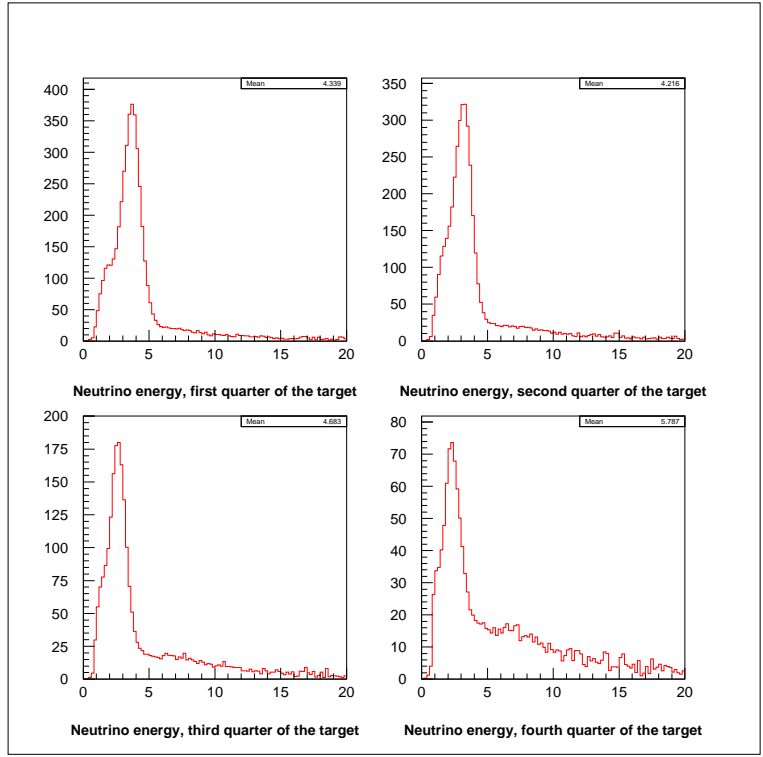,height=10cm}} 
\caption{Neutrino spectra from decays of particles produced in different sections
of the target.}
\label{targ_spect}
\end{figure}

\end{itemize} 

\begin{figure}[h]
\centerline{\epsfig{file= 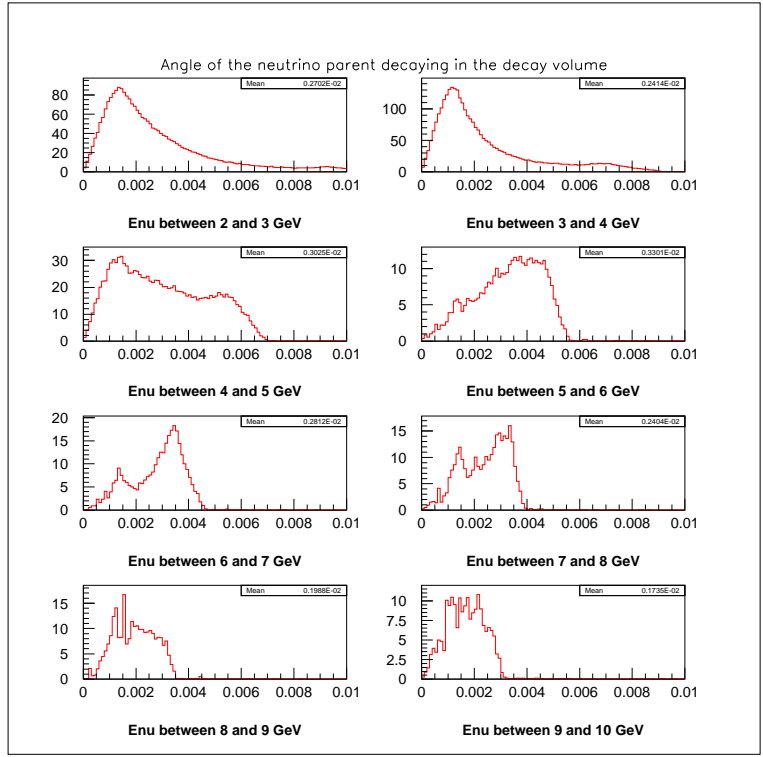,height=10cm}} 
\caption{Pion angle at the decay point for different  pion (neutrino) 
energies observed  in the near detector.}
\label{ang_diff_energ}
\end{figure}

The long production target in conjunction with the two-horn focusing system 
will produce the pion/kaon (neutrino) beam with characteristics dependent upon
the hadron energy.  Fig.\ref{ang_diff_energ} shows the distribution of the
angle of the neutrino parents after the focusing system. Low energy part of 
the spectrum comes from well focused beam. Neutrinos in the range 5 to 7 Gev
are produced by pions which are increasingly less and less focused, whereas
the high energy tail $E_{\nu} \geq 8 \:GeV$ are produced from pions produced
at very low angles and passing through the opening of the magnetic horns.  

\subsection{Decay Volume}
 
Pion and kaon beam formed by the horns decays in the decay volume consisting
of the last section of the target cave and the decay pipe. Neutrinos are
produced in two-body decay $\pi^{+} \rightarrow \mu^{+} + \nu_{\mu}$.
Energy and the flux of the produced neutrino depend on the decay 
angle $\theta_{dec}$ as:
\begin{equation}
  E_\nu=\frac{0.43E_{\pi}}{1+\gamma^{2}\theta^{2}}  
\end{equation}

\begin{equation}
Flux= \left(  \frac{2\gamma}{1+\gamma^{2}\theta^{2}} \right) ^{2}
\frac{A}{4\pi z^2}
\end{equation}

where 
\begin{itemize}
\item $\gamma=\frac{E_\pi}{m_{\pi}}$ is the Loretnz boost factor of a pion
\item $\theta$  is a decay angle, i.e. the angle between the pion and the
produced neutrino directions
\item $A$ is the area of the detector and $z$ is the distance between the decay
point and the detector.
\end{itemize}
 
\begin{figure}[h]
\hspace{1cm}\epsfig{file=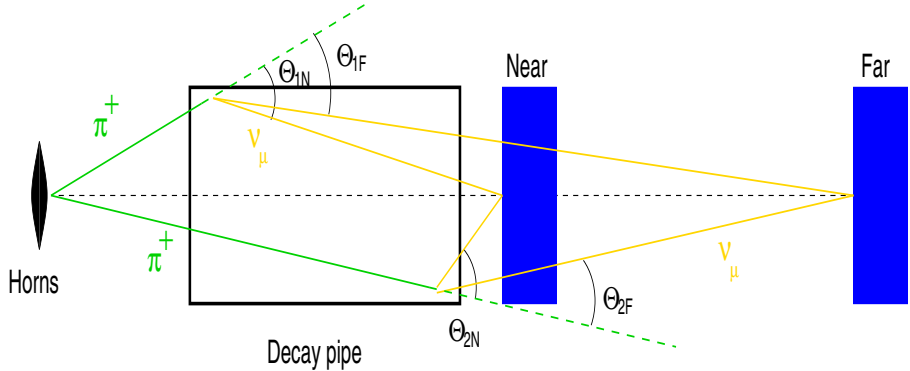,height=5cm,width=12cm}
\caption{Differences in the decay angles to reach the near and the 
far detectors due to finite decay pipe dimensions. A systematic difference 
$\Delta\Theta=\Theta_{N}-\Theta_{F}$ is increases with the $z$ of the
decay point. }
\label{decay_angles_far_near}
\end{figure}

 Finite transverse size of the decay volume, while important to maximize
the overall neutrino flux, causes a systematic difference between the
neutrino spectra observed at the near and far detectors. This is due to the
fact that the decay angle (i.e. the angle between the parent pion and 
resulting neutrino directions) necessary to reach the near and the far
detectors are different for decays occurring at finite radii as shown 
in Fig.\ref{decay_angles_far_near}. (A necessary
element of a rigorous proof is the approximate azimuthal symmetry of the
hadron beam and the fact that $\frac{dr}{dz}>\,0$ for the majority of the
focused particles.)
 
\begin{figure}[h]
\centerline{\epsfig{file=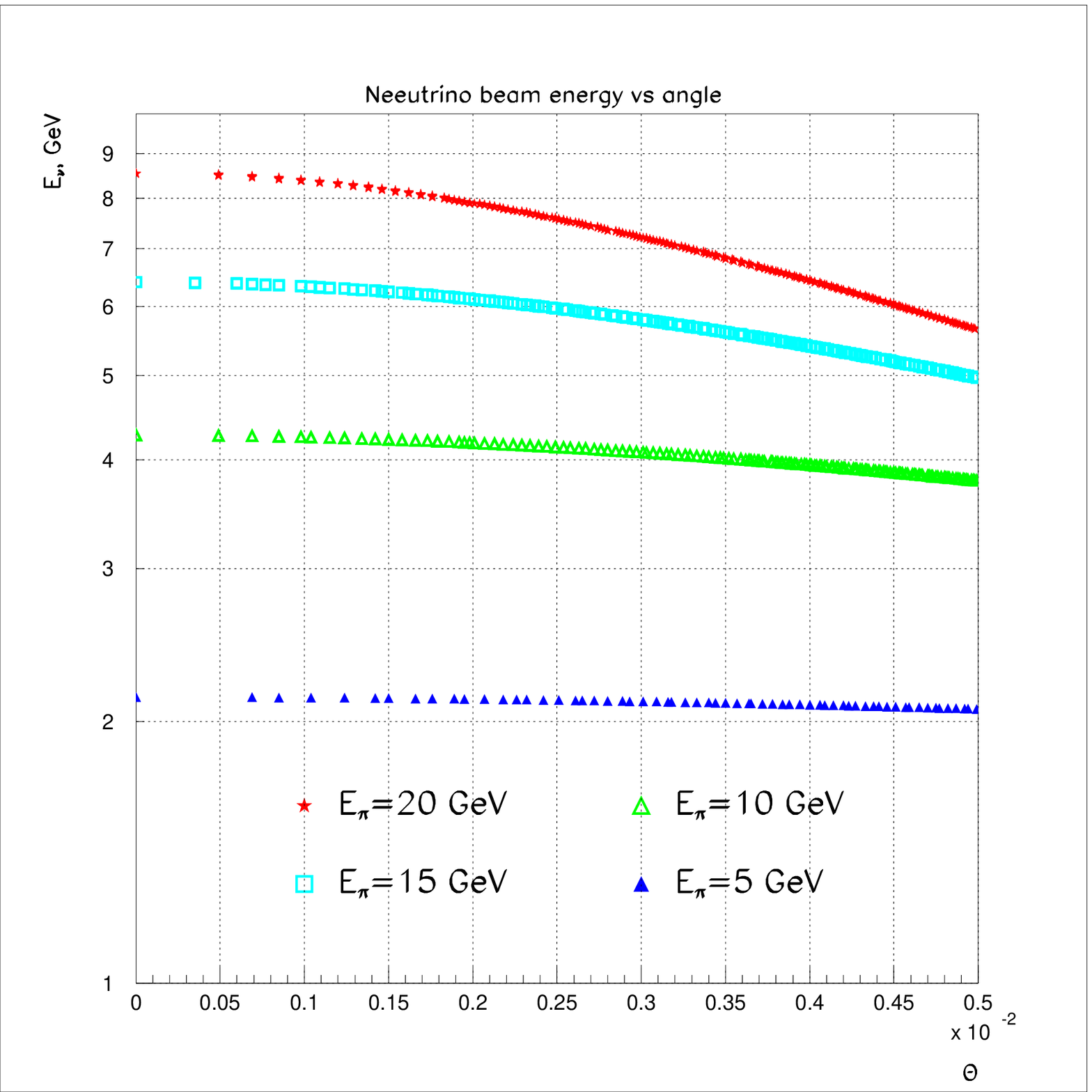,height=8cm}} 
\caption{Neutrino energy from the  $\pi^{+} \rightarrow \mu^{+} + \nu_{\mu}$
decay as a function of the decay angle for different pion energies.}
\label{enrg_vs_ang}
\end{figure}

\begin{figure}[h]
\centerline{\epsfig{file=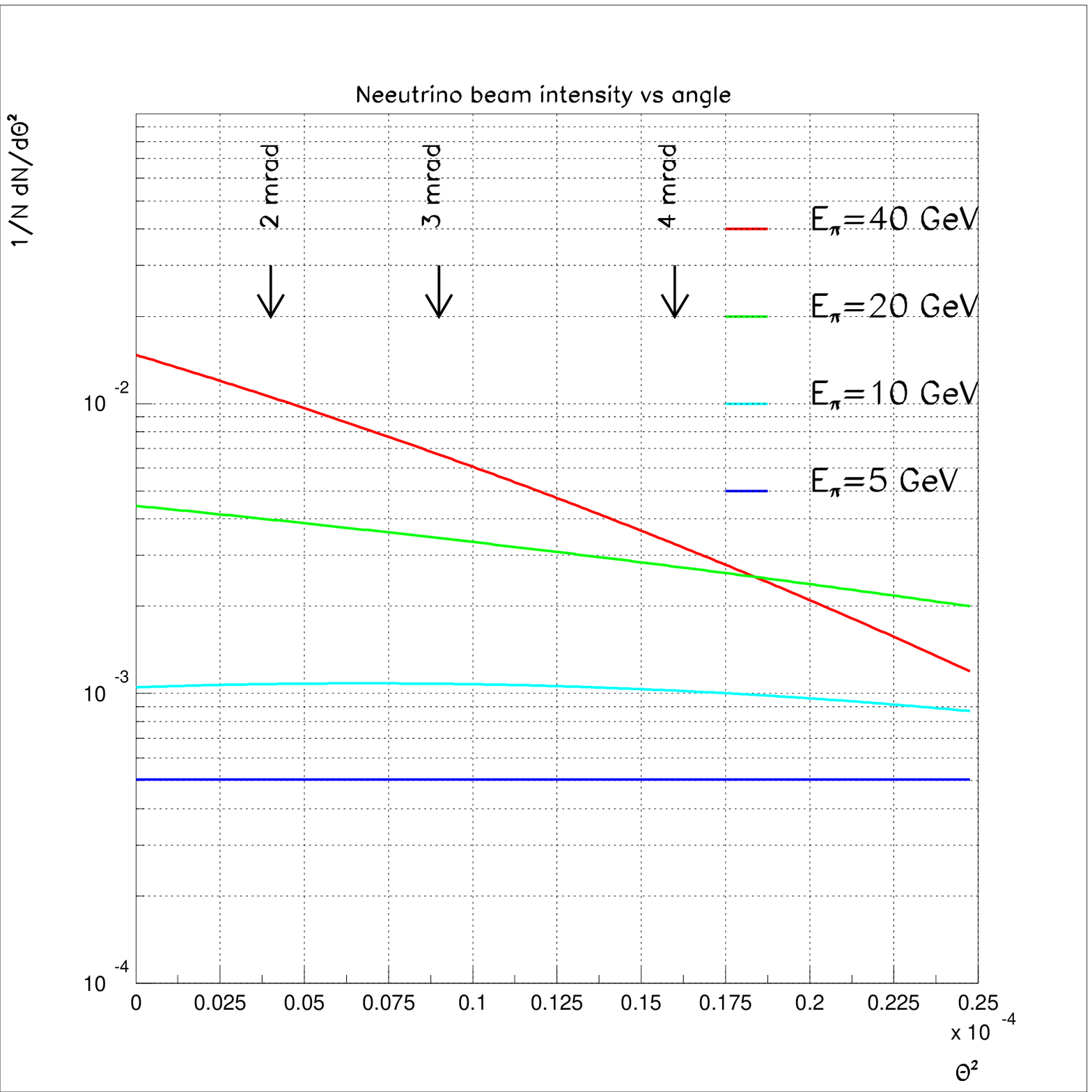,height=8cm}} 
\caption{Neutrino flux from the  $\pi^{+} \rightarrow \mu^{+} + \nu_{\mu}$
decay as a function of the decay angle for different pion energies.}
\label{flux_vs_ang}
\end{figure}

 Fig.\ref{enrg_vs_ang} illustrates the dependence
of the produced neutrinos and  Fig.\ref{flux_vs_ang} shows the neutrino
flux as a function of the decay angle.
This dependence is very strong for high energy pions, say 
$E_{\pi} \geq \:20\:GeV$. At the energies relevant to the main component
of the low energy beam, $E_{\pi} \leq \:10\:GeV$ neutrino energy is very weakly
dependent on the decay angle for $\theta_{dec} \leq 2 \: mrad$.      

\begin{figure}[h]
\centerline{\epsfig{file=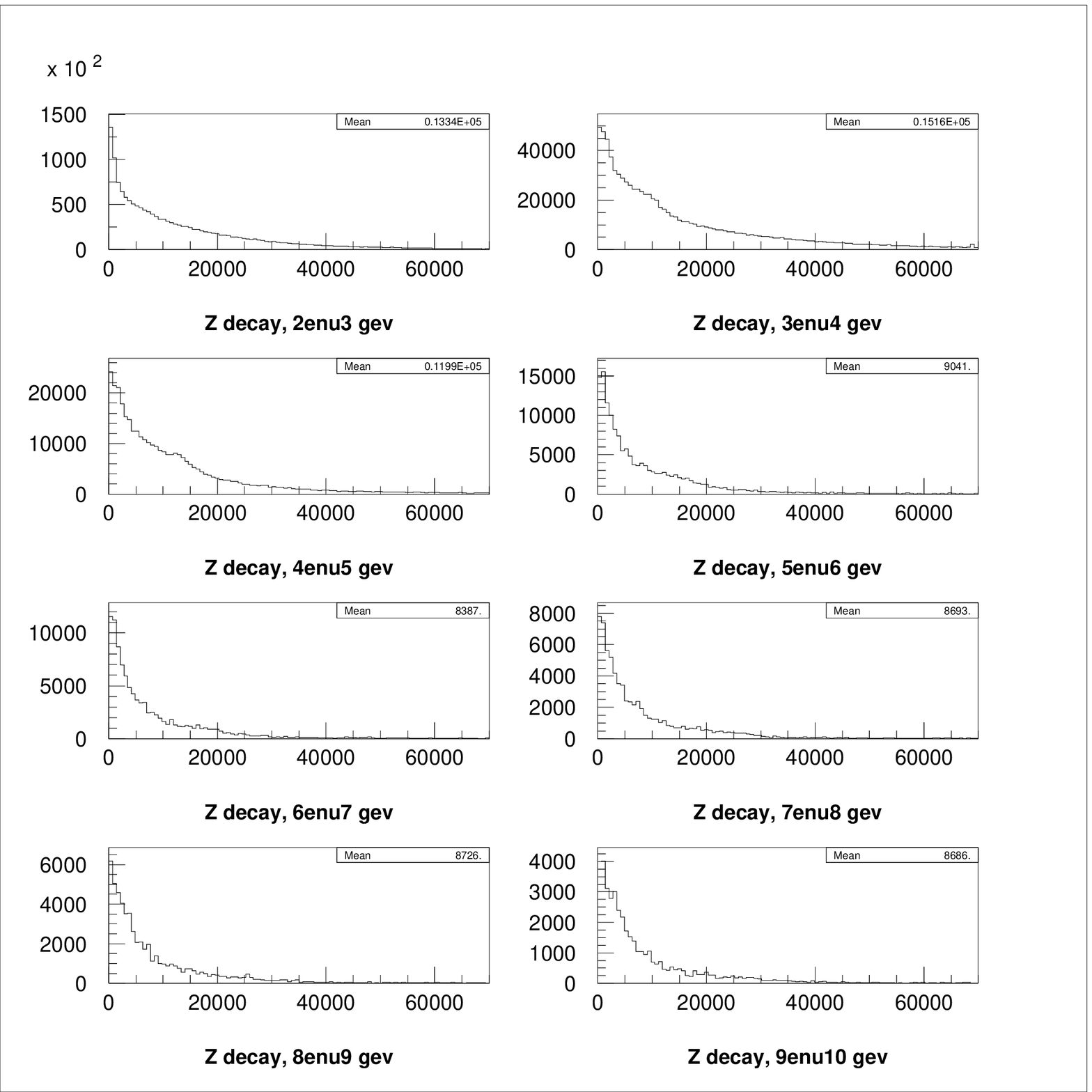,height=13cm}} 
\caption{Decay point of the neutrino parent for different  neutrino
energies.}

\label{z_vs_enrg}
\end{figure}

Hadron beam produced by the focusing system has an angular divergence of the
order of few miliradians. Such a divergence  produces a hadron beam much
larger than the radius of the decay volume, therefore most of the particles
will hit the walls of the decay pipe, unless they decay earlier.
Dependence of the angular distribution of pions on the pion energy (Fig.
\ref{ang_diff_energ}) will therefore lead to a  correlation between  the 
resulting neutrino energy and the average decay point of the parent pions
as shown in  Fig.\ref{z_vs_enrg}. This effect is contributing to a 
difference between neutrino spectra observed with the near and far detectors.
The neutrino flux observed by the detector varies with as $\Phi(z) \sim \frac
{\Phi_{0}}{(z_{det}-z_{dec})^2}$, where $z_{dec}$ is the average position
of the decaying pion. This reduction of the flux due to the angular 
divergence is independent of the neutrino energy, as the variation
of $z_{dec}$ is negligible compared to $(z_{det}-z_{dec})$. In case of the near
detector, though, the variation of of the attenuation of the flux with
the neutrino energy is significant.

\begin{figure}[h]
\centerline{\epsfig{file=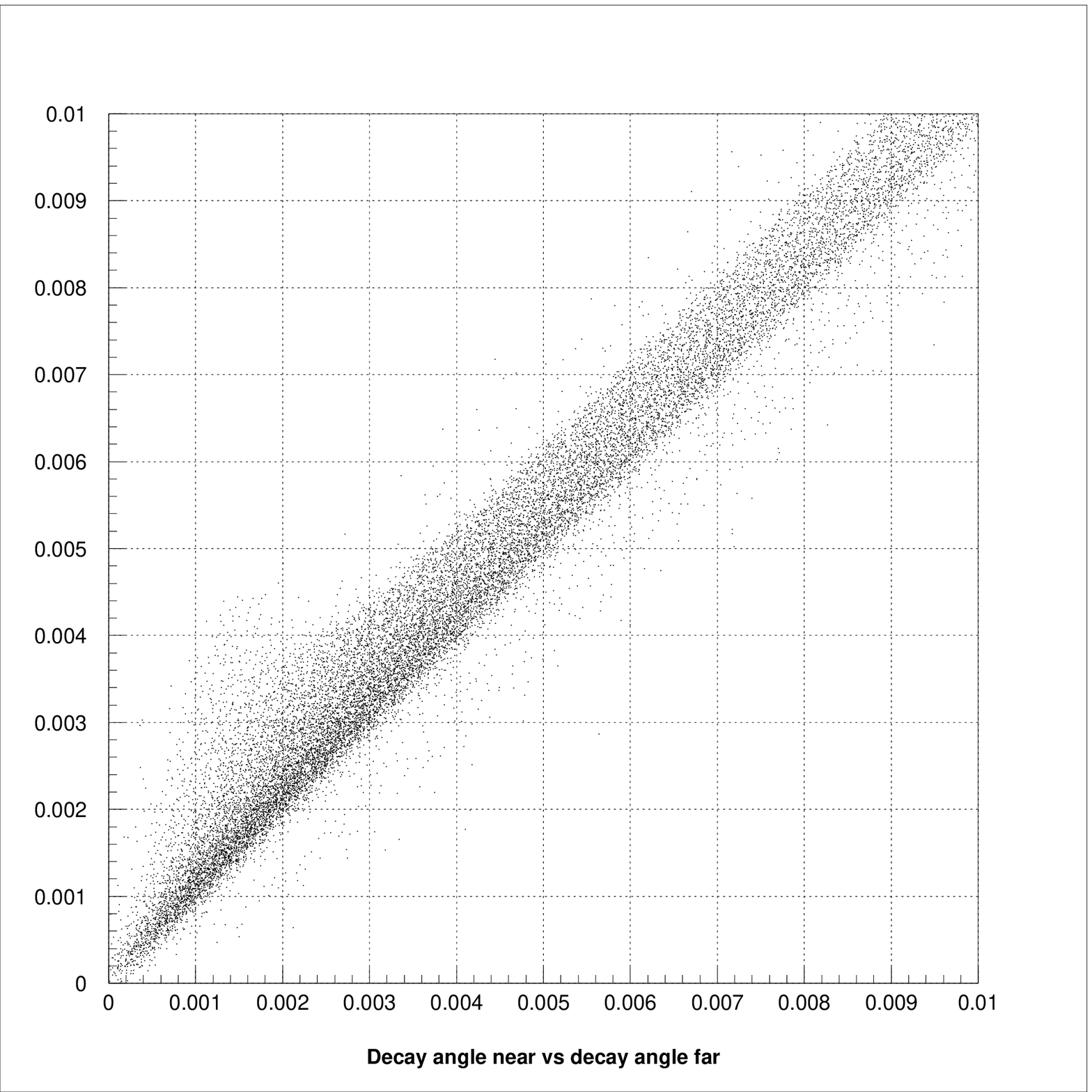,height=13cm}} 
\caption{Decay angle to reach the near detector vs the one to reach the far
detector. Spread of the angles is determined primarily the the transverse size
of the decay volume.}
\label{ang_far_near}
\end{figure}

Finite transverse size of the decay volume, in conjunction with the angular 
divergence  of the pion beam leads to another contribution between the
neutrino spectra observed in the near and far detectors. For a pion decaying
at some radius R inside the decay volume the decay angle pointing to
the near detector is larger than the decay angle necessary for neutrino
to reach the far detector, see Fig.\ref{ang_far_near}.   

\begin{figure}[h]
\centerline{\epsfig{file=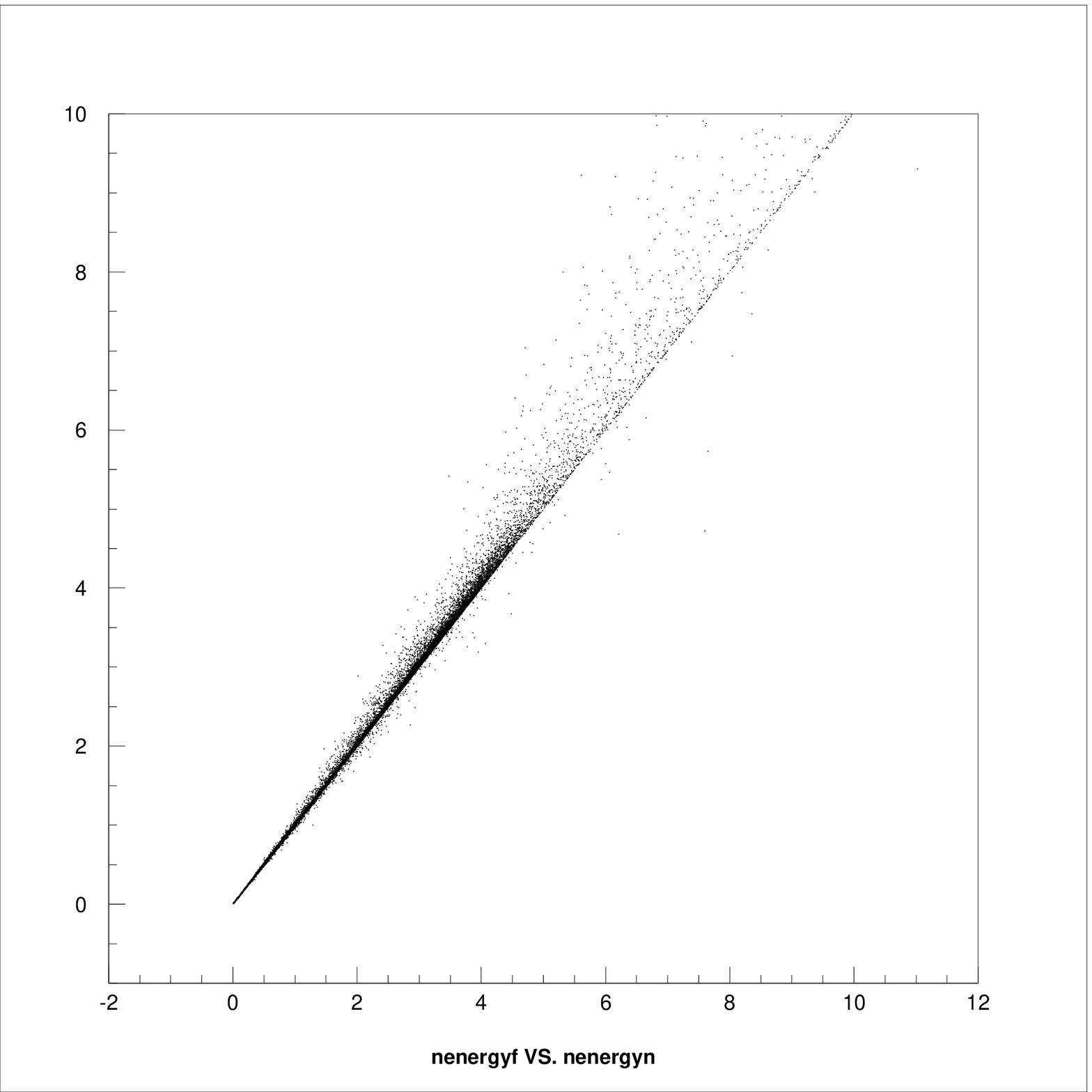,height=10cm}} 
\caption{Energy of the neutrino observed in the far detector vs the energy
observed in the near detector.}
\label{en_far_near}
\end{figure}

As a consequence of the dependence of the neutrino energy on the decay angle 
(Fig.\ref{enrg_vs_ang}) the neutrino energies detected at the near detector
will by systematicly lower in the near detector even if both neutrino fluxes
are produced by the same parent hadron beam, Fig.\ref{en_far_near}.

\subsection{Near and far neutrino fluxes}

Hadron beam produced at the target and focused by the horns gives rise to
the neutrino beam. Neutrino fluxes detected by the near and far detectors are
highly correlated, as they are produced by the same hadron beam. They are not 
identical though. The difference of the spectra at the far and near detectors
are primarily due to two effects:
\begin{itemize}
\item  
a solid angle subtended by the unit area at the near detector varies
considerably between the beginning and the end of the decay volume. This
effect is very small for the near detector. This effect is shown in Figs.\ref
{z_vs_far_near_1} and \ref{z_vs_far_near_2}

\begin{figure}[h]
\centerline{\epsfig{file=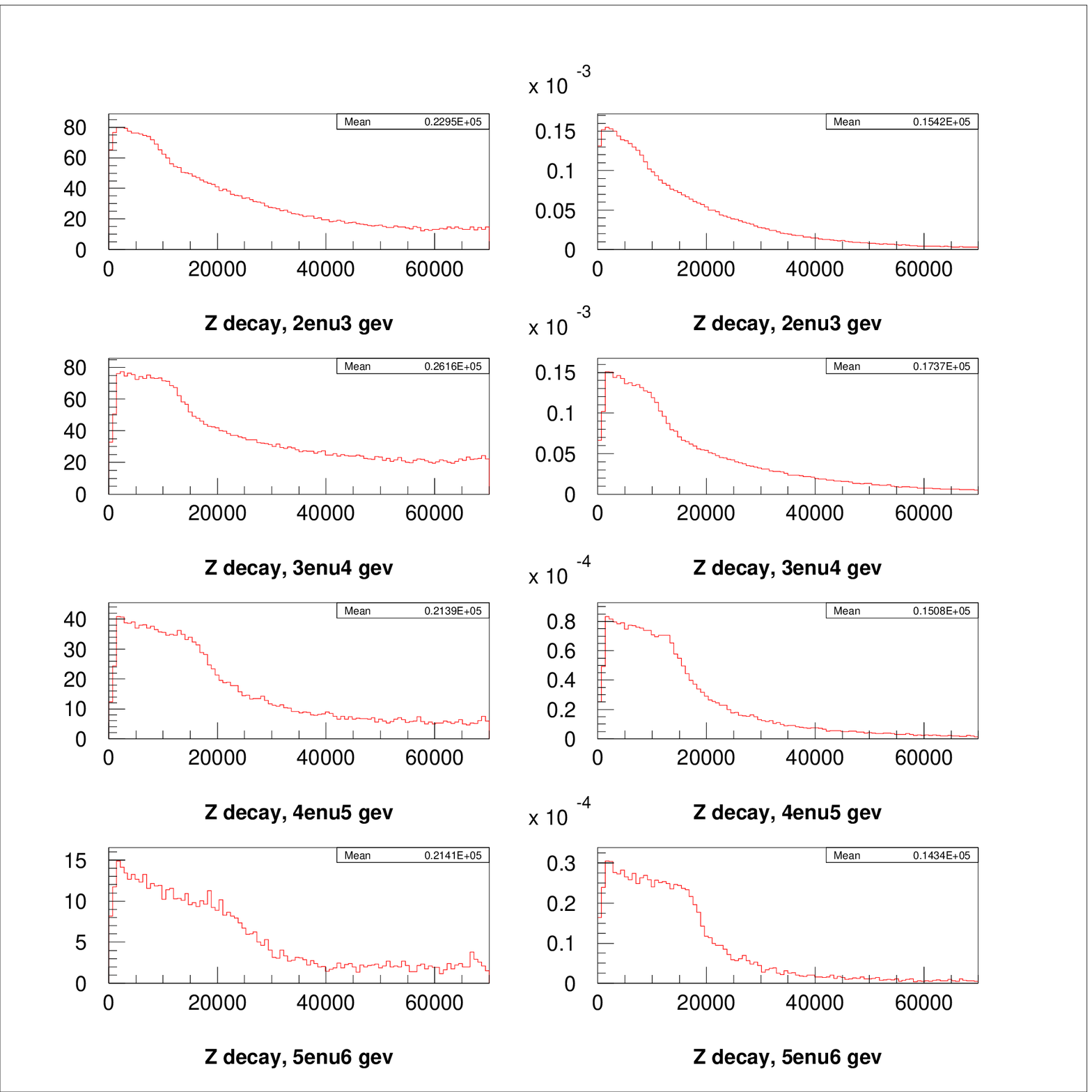,height=12cm}} 
\caption{Decay point of the neutrino parent for different  neutrino
energies. On the left are decays yielding neutrino in the near detector,
on the right there are decays yielding neutrino in the far detector}
\label{z_vs_far_near_1}
\end{figure}

\begin{figure}[h]
\centerline{\epsfig{file=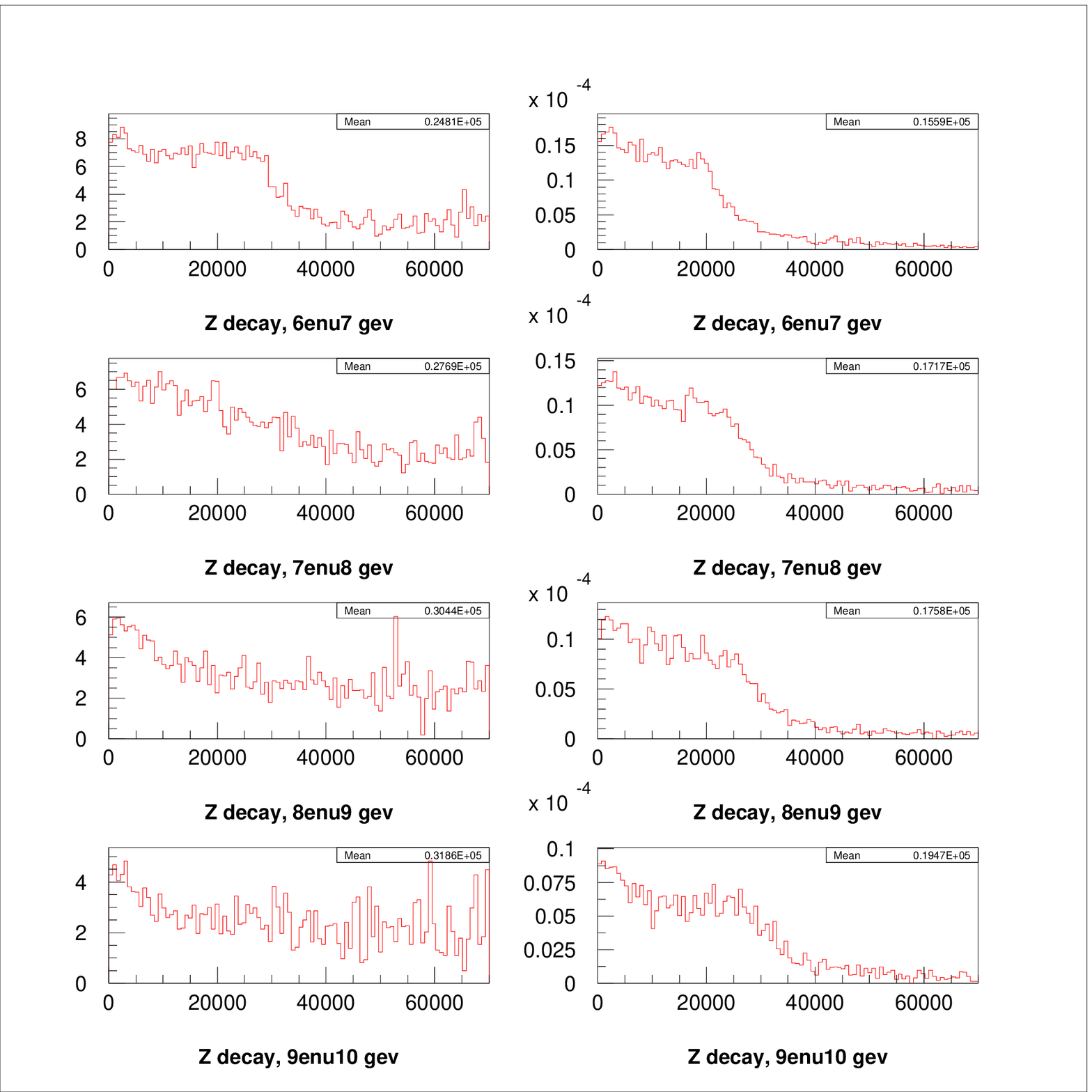,height=12cm}} 
\caption{Decay point of the neutrino parent for different  neutrino
energies. On the left are decays yielding neutrino in the near detector,
on the right there are decays yielding neutrino in the far detector}
\label{z_vs_far_near_2}
\end{figure}

As the result, the neutrinos produced towards the end of the decay pipe 
constitute much larger fraction of the neutrino flux detected at the 
near detector in comparison with the far detector. As the distribution
of the decay points along the decay pipe varies with the hadron momentum, 
the resulting energy distributions at both detectors will be somewhat 
different, leading to a variation of the ratio $R_{F/N}=\frac{dN^\nu_{F}}{dE}/
\frac{dN^{\nu}_{N}}{dE}$ as a function of neutrino energy. Thus the prediction of the far
detector spectrum from the spectrum measured at the near detector relies
on the proper modeling of the longitudinal distribution of the decay points.
The shape of longitudinal distribution of decay points is primarily
determined by the width of the decay volume and the angular divergence of
the hadron beam. At low hadron momenta a finite lifetime of pions and kaons
has a significant contribution.

\item  angular divergence of the hadron beam spreads particles throughout
the volume of the decay pipe. Neutrinos from decays at large radii will 
be produced at different decay angles to reach the near and the far detectors.
As the result, the energy of neutrinos observed at these two detectors will differ, 
with the energy at the far detector being in general higher, as shown in Fig.\ref
{en_far_near_vs_rad}.

\begin{figure}[h]
\centerline{\epsfig{file=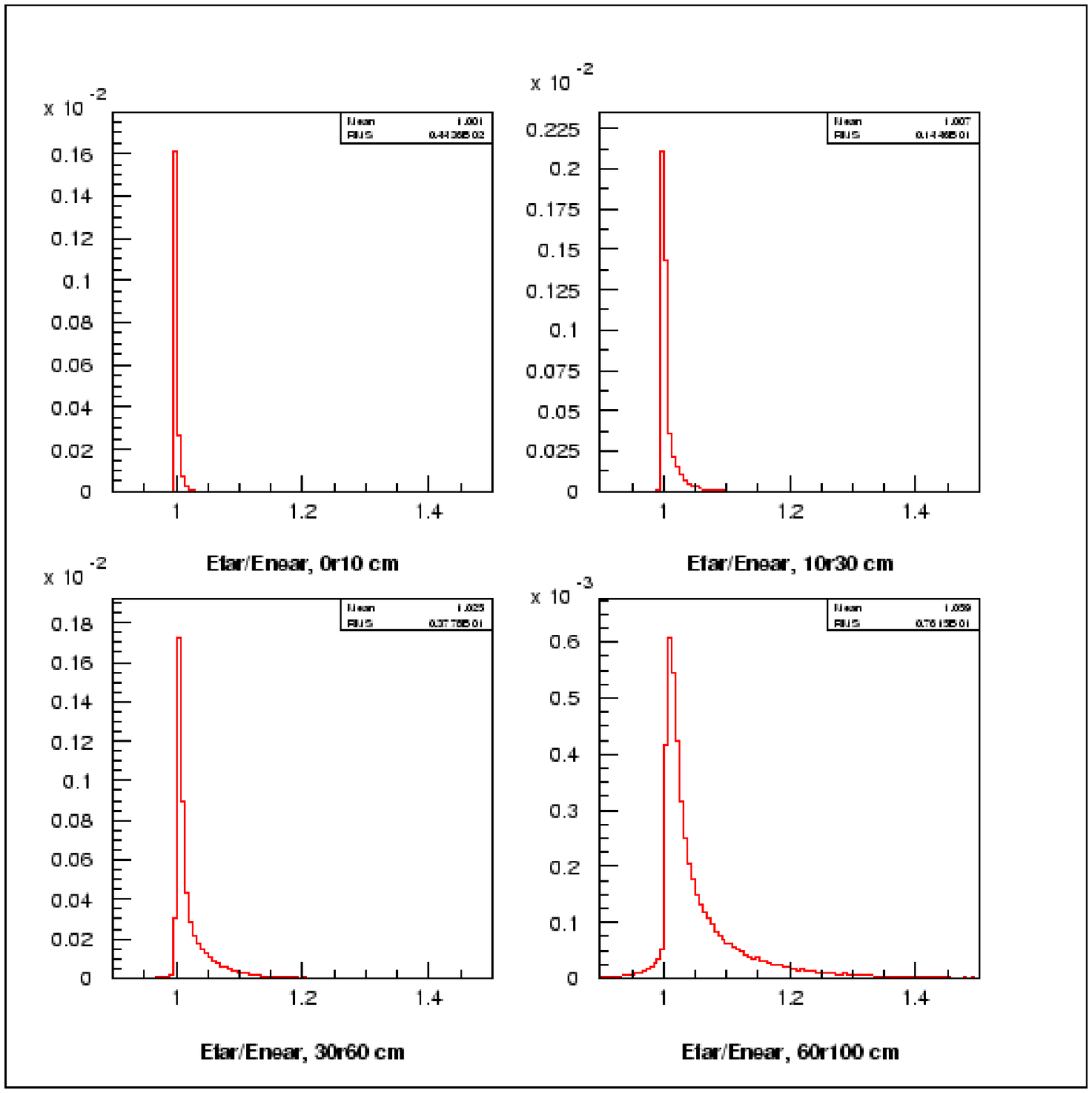,height=10cm}} 
\caption{Ratio of the energy observed in the far and near detectors
for neutrinos produced at different radii in the decay volume} 
\label{en_far_near_vs_rad}
\end{figure}

\end{itemize}

\section{A method to derive the Far Detector spectrum}

A neutrino spectrum observed at the far(near) detector can be derived
from the knowledge of the secondary hadron distribution (after the
focusing elements) and the geometry of the decay volume:

\begin{equation}
\frac{dN^{\nu}_{near}}{dE_{near}}=\iiiint F_{\pi/K}(r_{i},\theta,p)P_{\pi/K}(r_{i},
\theta,p,z)W_{\pi/K}(z,r_{dec},\theta,p,z_{near},E_{near})\,dr_{i}\,d\theta
\,dp\,dz
\label{near_spectrum}
\end{equation}

and
\begin{equation}
\frac{dN^{\nu}_{far}}{dE_{far}}=\iiiint F_{\pi/K}(r_{i},\theta,p)P_{\pi/K}(r_{i},
\theta,p,z)W_{\pi/K}(z,r_{dec},\theta,p,z_{far},E_{far})\,dr_{i}\,d\theta
\,dp\,dz
\label{far_spectrum}
\end{equation}

where:
\begin{itemize}
\item  $F_{\pi/K}(r_{i},\theta,p)$
 is the radial, angular ($\theta=\sqrt{p_{x}^{2}+p_{y}^{2}}/p_{z}$) and momentum
distribution of pions(kaons) after the focusing elements. This function is a 
convolution of the production cross section, horn acceptance and 
horn focusing. 
\item $P_{\pi/K}(r_{i},\theta,p,z)$
 is a probability that a pion(kaon)
with momentum $p$, radial position $r_{i}$ and the angle $\theta$ will
decay at the position $z$ along the decay volume. The radial position of the
decay will be $r_{dec} = r_{i} + \theta z$. For simplicity we have set
$z=0$ at the end of the focusing system. This function depends solely on
the geometry of the decay volume
\item  

$W_{\pi/K}(z,r_{dec},p,\theta,z_{near/far},E_{near/far})$
is a probability that
a pion(kaon) with momentum $p$ and the angle $\theta$ decaying at the position $z$ 
along the decay volume at the radius $r_{dec}$ will produce neutrino with
energy $E_{near/far}$ at the center of the respective detector. This function
is determined purely by two-body kinematics. 
\end{itemize}  

For simplicity of the presentation we have assumed an axial symmetry of
the problem. In practice the integral is over a  six-dimensional space
of $(x,y,p_{x},p_{y},p_{z},z)$. A complete treatment must also include neutrinos
produced before the end of the focusing system.

\subsection{First Approximation: Pencil-like Beam}

In case of a pencil-like hadron beam, i.e. a beam with no angular divergence,
directed towards the far and near detectors the relation between the near
and far neutrino spectra is simplified. The decay angle $\theta=0$ and the
neutrinos detected at both detectors have the same energy 
$E_{near}=E_{far}=0.43E_{\pi}$. Thus the measured neutrino flux at the near 
detector provides
a direct measurement of the energy distribution of the parent pion beam.

The neutrino spectra observed at the near and far detectors are not identical,
though. They differ, because of the differences in the acceptances of the
near and far detector. The acceptance is a function of the distance from
the decay point to the detector. This distance is practically the same for
all neutrino energies in the case of the far detector, but it is a function
of energy, due to the finite pion life-time, at the near detector position.  
Thus the neutrino spectrum at the far detector can be derived from the
spectrum observed at the near detector as:

\begin{equation}
\frac{dN_{far}}{dE} = \left[\frac{\frac{dN_{far}}{dE}}{\frac{dN_{near}}{dE}}\right]_{MC}
\frac{dN_{near}}{dE}= T(E)\frac{dN_{near}}{dE}
\end{equation}

The transfer function $T(E)$, in this simple case, is in fact calculable from
the known pion life-time, the length of the decay region and the position
of the near and far detectors.

\subsection{Second Approximation: Small Aperture Beam and  'Double Ratio' Method}

\begin{figure}[h]
\centerline{\epsfig{file=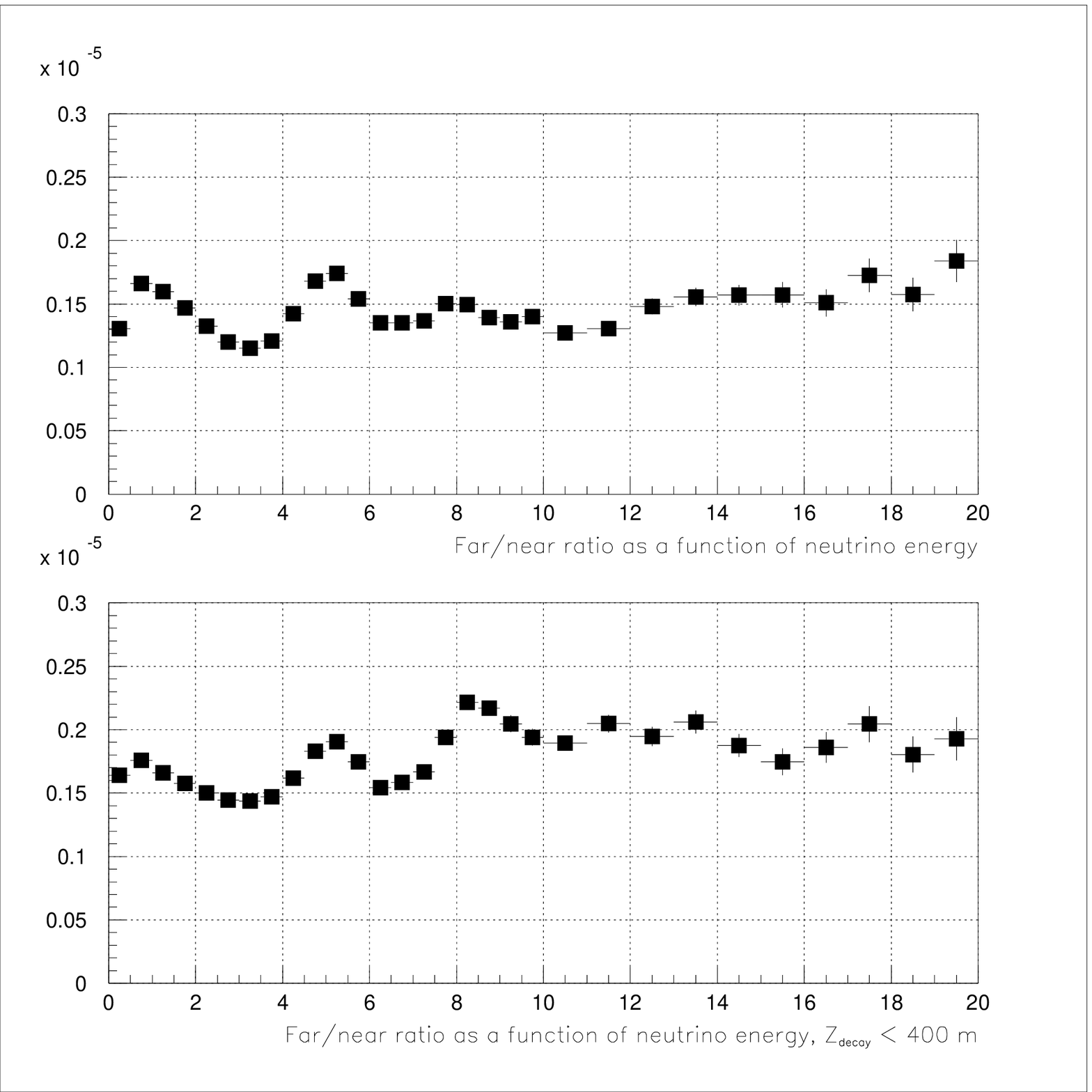,height=10cm}} 
\caption{Ratio of the neutrino fluxes observed in the far and near detectors
as a function the neutrino energy. Top figure is for all neutrinos, bottom figure
is for neutrinos originating in the first section of the decay volume, $z_{decay}<400 m$} 
\label{far_over_near}
\end{figure}

Focusing of  particles with large momentum spread and produced  in an 
extended target into a parallel beam is not
possible. A realistic beam has a very sizeable angular divergence ( see
Fig.\ref{ang_diff_energ}) which is different for pions of different energies.
This effect determines the effective distribution of the decay points
along the length of the decay volume, as shown in Figs.\ref {z_vs_far_near_1}
and \ref{z_vs_far_near_2}. Derivation of the transfer function $T(E)$
requires, in this case, a Monte Carlo simulation of the production and 
focusing of the pions and kaons. The transfer function  $T(E)$ is just a ratio
of integrals \ref{far_spectrum} and \ref{near_spectrum} and its shape 
is shown in Fig.\ref{far_over_near}
The mean value of the function, $\sim 0.15\times 10^{-5}$, is given by the square
of ratio of the far and near detector distances from the neutrino source.
The shape of this function is related to the variation of the longitudinal distribution
of the decay points along the beam line as shown in Figs.\ref{z_vs_far_near_1} 
and \ref{z_vs_far_near_2}. The structures observed in this function are reflections of
the acceptance of the focusing elements. A sharp rise around $5\,GeV$ is a related 
to the fact that the pions with $E_\pi \geq 12\,GeV$ start missing the first horn
and loose benefit of the focusing. As a result their angular distribution broadens,
and they hit walls of the decay volume before they decay. The average distance
from the decay point to the near detector increases (in comparison with the focused
pions of lower energy) and the flux observed at the near detector is reduced
thus producing a rise in the $Far/Near$ ratio. An analogous mechanism, involving
the horn 2 is responsible for the rise of the $Far/Near$ ratio around $8\,GeV$. 

The transfer function will depend, somewhat, on the assumed production spectra in 
the target, but its overall value and the detailed shape is chiefly determined
by the geometry of the beam line. 

For a beam of a very small radial aperture, such that decay angles leading to the
near and far detectors are very similar, the energy of neutrinos observed
at the both detectors will be very close (see Fig.\ref{en_far_near_vs_rad})
and the dependence on the production cross sections will cancel, to a first
order, in a ratio $ \left({\frac{dN_{far}}{dE}}/{\frac{dN_{near}}{dE}}\right)_{MC}.$

This method of predicting the far detector spectrum, often referred to as
a `double ratio' method is widely used to estimate the systematic errors
in the disappearance experiments.  Owing to a smooth behavior of the underlying
production cross sections, acceptances and focusing, this is a quite robust
method of predicting the far detector flux. 

Shortcoming of this method comes to light when some instrumental effects modify
the neutrino (or pion) spectra in a very limited energy range. An example of
such an effect is a small displacement of focusing elements (horns) whereby
some small part of the phase space of the produced particles becomes focused
and produces neutrinos of some energy, at the expense of particles in a neighboring part
of the phase space, being  moved out of the horns acceptance and hence producing
significantly fewer neutrinos. As a result of the extended transverse size of
the decay volume these 'gained' and 'lost' neutrinos will have somewhat different
energies in the near and far detectors, thus leading to a characteristic bipolar
shape of the predicted far detector spectrum fluctuating around the real 
observed spectrum. 

The limitation of the `double ratio' method becomes apparent when one attempts
to predict a spectrum of the far detector in the 'off-axis' position, where
the spectrum of the observed neutrinos is completely different then that in the
near detector.

\subsection{A Far-to-Near Correlation Matrix Method}

Neutrino spectra in the far and near detectors are correlated, as they 
result from the decays of the same parent hadron beam. Thus, every neutrino
with energy $E_{n}$ observed at the near detector implies a certain flux
of neutrinos with energy distribution $E_{f}$ given by
\begin{equation}
\frac{dN}{dE_{f}} =\frac{W(z,r_{dec},p,z_{f},E_{f})}{W(z,r_{dec},p,z_{n},E_{n})}
\end{equation}
This flux, implied for the far detector, depends on the position of the decay point
as well as on the momentum vector of the parent hadron.
Integration over all decays yields a matrix, which correlates the spectra
at both locations.  The matrix element 
$M_{\Delta E_{f}\Delta E_{n}}$ is given by 
\begin{equation}
M_{\Delta E_{f}\Delta E_{n}} = \frac{
\idotsint F_{\pi}(r_{i},\theta,p)P(r_{i},
\theta,p,z)W(z,r_{dec},\theta,p,z_{f},E_{f})\,dr_{i}\,d\theta
\,dp\,dz\,dE_{f}}
{\idotsint F_{\pi}(r_{i},\theta,p)P(r_{i},
\theta,p,z)W(z,r_{dec},\theta,p,z_{n},E_{n})\,dr_{i}\,d\theta
\,dp\,dz,dE_{n}}
\label{matrix}
\end{equation}

where the integrals are over the bin sizes $\Delta E_{f}$ and $Delta E_{n}$,
respectively.
In order to relate the observed event spectra (rather than the neutrino spectra) at
both detectors the weight functions $W(z,r_{dec},\theta,p,z_{n/f},E_{n/f})$
should be replaced in the integral (\ref{matrix}) by 
$W(z,r_{dec},\theta,p,z_{n/f},E_{n/f})\times\sigma_{tot}^{\nu_\mu}(E_{n/f})$.
  
Given a vector containing the observed event spectrum in the near detector,
$N = \left( N_{1},N_{2},\cdots,N_{n} \right)$ the predicted far detector
spectrum $F = \left( F_{1},F_{2},\cdots,F_{n} \right)$ can be derived as:

\begin{equation}
\left( F_{1},F_{2},\cdots,F_{n} \right) = 
\begin{bmatrix}
M_{11}&M_{12}&\dots&M_{1n}\\
M_{21}&M_{22}&\dots&M_{2n}\\
\hdotsfor{4}\\
M_{n1}&M_{n2}&\dots&M_{nn}\\
\end{bmatrix}
\begin{bmatrix}
N_{1}\\
N_{2}\\
\hdotsfor{1}\\
N_{n}
\end{bmatrix}
\label{far_prediction}
\end{equation}

We have computed the correlation matrix $M$ for the event spectra in $0.5\,GeV$ bins  
using a set of Ntuples generated using GNUMI simulation program and corresponding to $1.7\times10^7$ protons on
target \cite{gokhan}. Uppermost $15\times15$ fragment of the matrix, thus showing the correlation
between far and near detector event spectra below $7.5\,GeV$  is shown in (\ref{M}). Columns of the matrix
are related to the energy observed in the far detector, rows to the energy in the near detector.

\begin{equation}
\hskip -2.0cm
\tiny
\setcounter{MaxMatrixCols}{20}
M=\begin{bmatrix}
 1.30 &   0  &   0  &   0  &   0  &   0  &   0  &   0  &   0  &   0  &   0  &   0  &   0  &   0  &   0  \\
 0.04 & 1.65 &   0  &   0  &   0  &   0  &   0  &   0  &   0  &   0  &   0  &   0  &   0  &   0  &   0  \\
   0  & 0.03 & 1.59 &   0  &   0  &   0  &   0  &   0  &   0  &   0  &   0  &   0  &   0  &   0  &   0  \\
   0  &   0  & 0.03 & 1.45 &   0  &   0  &   0  &   0  &   0  &   0  &   0  &   0  &   0  &   0  &   0  \\
   0  &   0  &   0  & 0.05 & 1.29 &   0  &   0  &   0  &   0  &   0  &   0  &   0  &   0  &   0  &   0  \\
   0  &   0  &   0  &   0  & 0.08 & 1.15 &   0  &   0  &   0  &   0  &   0  &   0  &   0  &   0  &   0  \\
   0  &   0  &   0  &   0  &   0  & 0.11 & 1.06 &   0  &   0  &   0  &   0  &   0  &   0  &   0  &   0 \\ 
   0  &   0  &   0  &   0  &   0  &   0  & 0.17 & 1.03 &   0  &   0  &   0  &   0  &   0  &   0  &   0  \\
   0  &   0  &   0  &   0  &   0  &   0  &   0  & 0.23 & 1.05 & 0.01 &   0  &   0  &   0  &   0  &   0  \\
   0  &   0  &   0  &   0  &   0  &   0  &   0  & 0.01 & 0.35 & 1.00 & 0.01 &   0  &   0  &   0  &   0  \\
   0  &   0  &   0  &   0  &   0  &   0  &   0  &   0  & 0.03 & 0.47 & 0.84 & 0.01 &   0  &   0  &   0  \\
   0  &   0  &   0  &   0  &   0  &   0  &   0  &   0  & 0.01 & 0.07 & 0.56 & 0.63 & 0.01 &   0  &   0  \\
   0  &   0  &   0  &   0  &   0  &   0  &   0  &   0  &   0  & 0.02 & 0.15 & 0.58 & 0.52 & 0.01 &   0  \\
   0  &   0  &   0  &   0  &   0  &   0  &   0  &   0  &   0  & 0.01 & 0.05 & 0.24 & 0.51 & 0.45 & 0.01 \\
   0  &   0  &   0  &   0  &   0  &   0  &   0  &   0  &   0  & 0.01 & 0.02 & 0.10 & 0.29 & 0.46 & 0.41
\end{bmatrix}\times 10^{-6}
\label{M}
\end{equation}

\setcounter{MaxMatrixCols}{10}

\normalsize

The matrix provides a very good representation how the far detector spectrum relates to the near one. Column $10$,
for example, states that $1$ event observed in the near detector in the energy bin between $4.5\,GeV$ and
$5\,GeV$ implies (modulo an overall factor $10^{-6}$) that the far detector should register
$1$ event with energy between  $4.5\,GeV$ and $5\,GeV$, $0.47$ events  between  $5.0\,GeV$ and $5.5\,GeV$,
 $0.07$ events  between  $5.5\,GeV$ and $6.0\,GeV$ etc. 
 
The  matrix is computed for
 all the simulated $\nu_{\mu}$ CC events, regardless of the actual
neutrino parent.  To a good approximation,
no background from interactions of other neutrino types (including
$\bar{\nu_{\mu}}$) needs to be included, as the detection of a $\mu^{-}$
uniquely identifies a $\nu_{\mu}$ charged current event.
For the present studies, smearing was  ignored, i.e., matrix
elements were calculated from generated near and far detector energies.

For the analysis purposes the correlation matrix $M$ should be computed
for the spectra with smearing and acceptance cuts properly included.

\section{Systematic Errors: Sensitivity of the Prediction to the
Details of the Parent Hadron Beam}

When the near-to-far correlation matrix $M$ is computed in the same
model as the near and the far detector spectra, the predicted  far 
spectrum using the equation \ref{far_prediction} matches {\it exactly}
the actual far spectrum. It implies that the far detector spectrum would be
known exactly, should the beam simulation program describe the real
beam in all details. In practice, the understanding of the beam is somewhat 
limited, hence leading to  systematic errors on the far spectrum prediction.
In general, these errors will be due to the differences of the spatial
and momentum distribution of pions decaying within the decay volume. 

\subsection{How Small the Systematic Errors Should be?}

Clearly, smaller the systematic errors are - the better. In practice
there is always a threshold below which a further reduction of the systematic
error of the experiment has no impact on the physics results.

A judgment of the impact of the systematic errors on the physics result is
fairly complicated. It depends on several aspects of the experiment, like:
\begin{itemize}
\item how large the measured effects are with respect to the systematic
errors?
\item what is a type of a measurement? For example a fairly small, but
systematic distortion of a spectrum may have a significant influence on the
determined shape parameters.
\item what are the experimental smearing effects?
\item what are the statistical errors?
\ etc...
\end{itemize}

Neutrino disappearance experiment, like MINOS, is expected to detect
and measure a very dramatic effect of $100\%$ disappearance at a particular
neutrino energy. It is very difficult to imagine systematic errors related to 
the understanding of the neutrino beam which could create false effects
of this magnitude or could significantly distort the observed effect
of full disappearance. 

On the other hand a detailed shape of a disappearance curve has a significant
physics interest. In the least interesting case it will provide yet another
proof of the underlying physics mechanism. In a more interesting case it
may provide a clue for physics phenomena beyond the current orthodox picture.

\begin{figure}[h]
\centerline{\epsfig{file=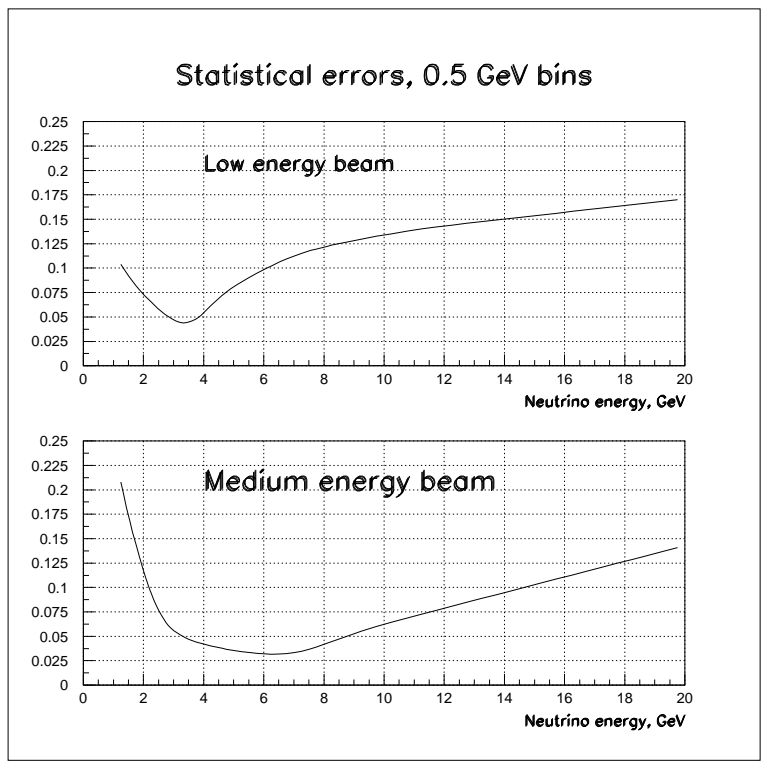,height=10cm}} 
\caption{Statistical errors in $0.5\,GeV$ bins of  far detector neutrino event
spectrum for $5\,kton$ detector
and exposure equivalent to $8\times10^{20}$ protons on target}
\label{stat}
\end{figure}

Statistical errors of the experiment provide a reasonable yardstick
for the systematics error. The expected statistical errors of the observed
neutrino flux at the $5\,kton$ far detector and an exposure corresponding
to $8\times10^{20}$ protons on NuMI target are shown in Fig.\ref{stat} for
the case of $0.5\,GeV$ bins. These error refer to the 'no-oscillation' case.

In reality the reduction of the statistics induced by the neutrino 
disappearance
as well as the experimental smearing effects will tend to further reduce the
importance of the systematic errors.  
 
\subsection{Dependence on the Particle Production Model}
\label{prod_model_dependence}
The extent, to which the distributions of particles in the integral 
(\ref{matrix}) defining the correlation matrix change with different 
production models,
and, as a result, the matrix elements change, contributes to the systematic 
error. 

Dependence of the predicted far detector flux can be evaluated by 
applying the same correlation matrix $M$ to the near detector spectra,
as derived in various production models, and comparing the predicted far
detector spectra computed from Eq.\ref{far_prediction} with the actual ones. 

There are no GEANT interfaces to the production models other than FLUKA.
Different production models can  be studied by re-weighting the neutrinos
according to the production cross section of its parent. Such a weighting
procedure, developed by Mark Messier\cite{MMessier}, exists only for the parent $\pi^{+}$.

\begin{figure}[h]
\centerline{\epsfig{file=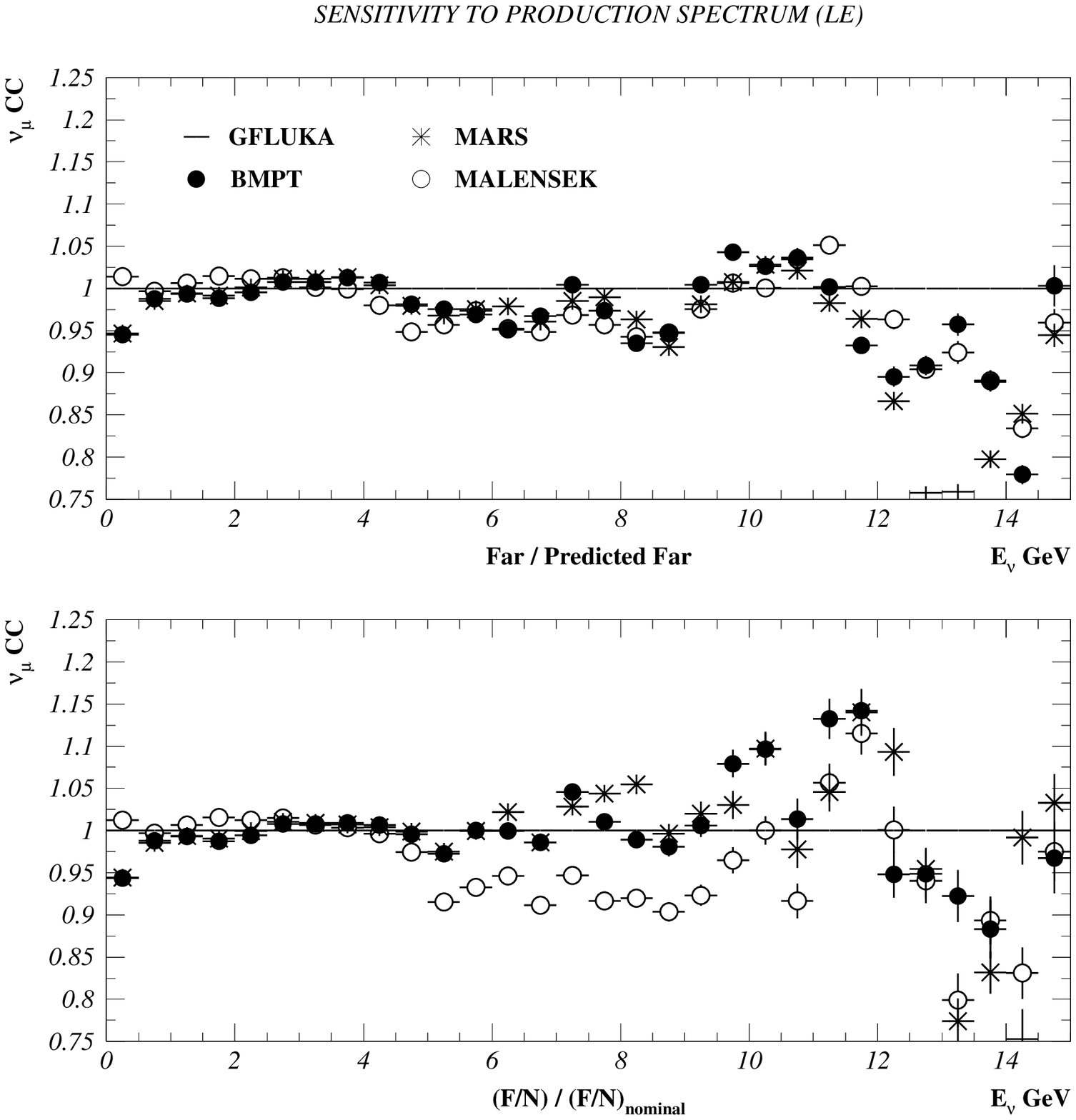,width=11cm}}
\caption{Upper figure: ratios of far detector spectra predicted by
various production
models  to those predicted from the corresponding near spectra using 
 the correlation   matrix $M$ calculated with the GFLUKA
model in ow energy beam. Lower figure: a
comparison of the Far/Near ratios in different prodution models over 
the nominal GFLUKA Far/Near ratio.}
\label{prod_mod_le}
\end{figure}

The correlation matrix $M$ was re-computed  for the
neutrinos produced by the pions only and leaving aside the neutrinos 
produced from kaon and muon decays as well as neutrinos produced in
proton-induced showers in  the final absorber. This matrix was subsequently
applied to the near detector spectrum, re-weighted for different models
and the results were compared with the far detector spectrum in these 
models. Results are shown in Fig.\ref{prod_mod_le}.

Far detector spectrum is predicted with the accuracy of the order of
$2\%$ for the component of the neutrino beam produced by the focused
part of the hadron beam. The prediction is accurate to $\sim 5\%$
for neutrinos energies up to $\sim12\,GeV$, the sensitivity of the method
to the production spectra somewhat reduced in comparison with the 
'double ratio' method.

\subsection{Production Model Dependence: Medium Energy Beam}

\begin{figure}[h]
\centerline{\epsfig{file=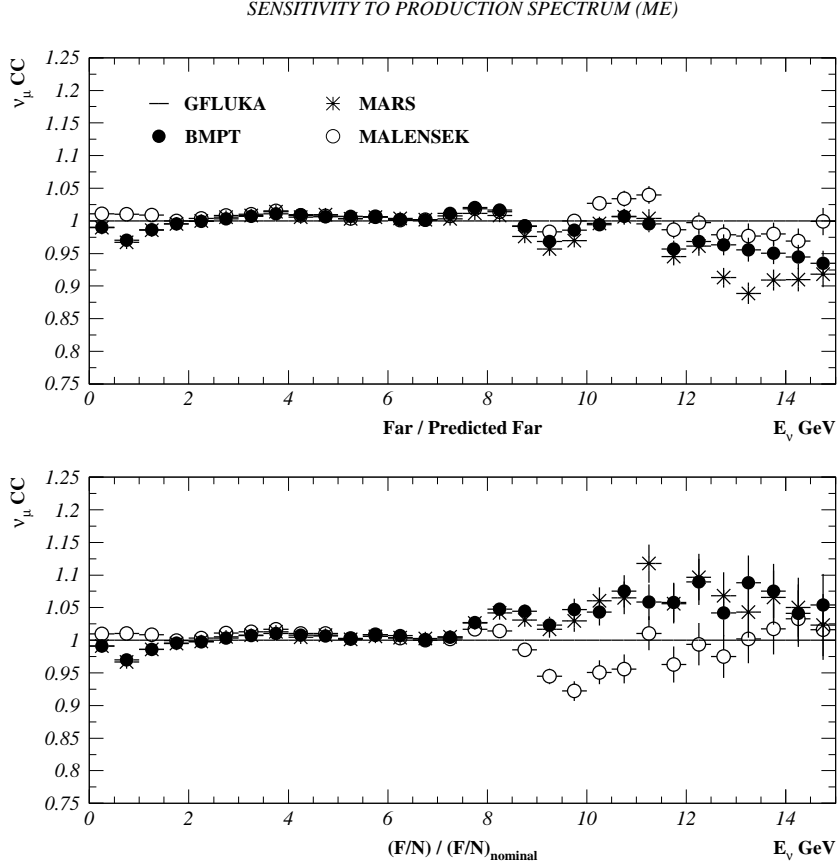,width=11cm}}
\caption{Upper figure: ratios of far detector spectra predicted by
various production
models  to those predicted from the corresponding near spectra using 
 the correlation   matrix $M$ calculated with the GFLUKA
model. Medium energy beam option. Lower figure: a
comparison of the Far/Near ratios in different prodution models over 
the nominal GFLUKA Far/Near ratio.}
\label{prod_model_me}
\end{figure}

The procedure described in the section\ref{prod_model_dependence}
was repeated for the case of the medium energy beam. The results
are shown in Fig.\ref{prod_model_me}.

As before, the prediction is good to $\sim2\%$ for the entire region 
of peak intensity and it remains good to $\sim 5\%$ for energies up to
 $\sim12\,GeV$. The sensitivity to the production model details is,
again, reduced in comparison with the 'double ratio' method.

\subsection{Sensitivity to the Details of  Modeling of the Neutrino Beam}

Functions $F_{\pi}(r_{i},\theta,p)$ and $P(r_{i},\theta,p,z)$
used to compute the correlation matrix matrix $M$ depend on the momentum
and spatial distribution of hadron in the decay volume, hence they depend
on the geometry and focusing characteristics of the beam forming
elements. How sensitive is the resulting matrix to the details of the
understanding of the beam line?

Two programs are currently in use to simulate the neutrino fluxes:
\begin{itemize}
\item GNUMI is a GEANT-based full simulation of the production and focusing
of particles. It includes scattering and re-interactions of the produced
hadrons. 
\item PBEAM is a much faster, albeit somewhat simplified, intended for 
optimization of the beam line design. 
\end{itemize}

\begin{figure}[h]
\hspace{1.5cm}\epsfig{file=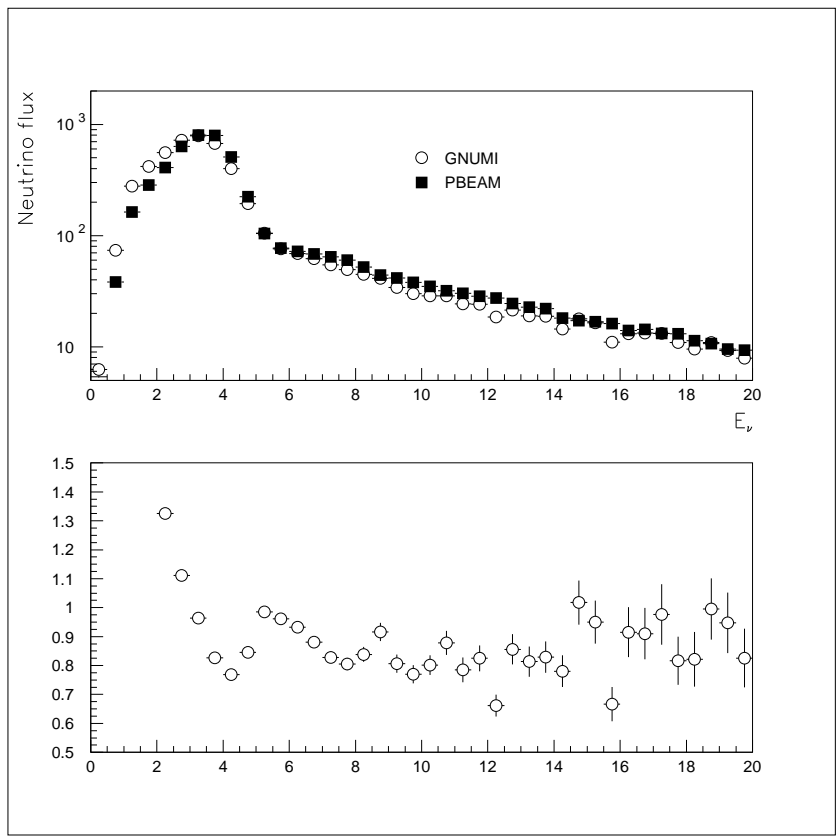,height=11cm}
\caption{Up: near detector spectra as predicted by PBEAM and GNUMI programs
Down: a ratio of the near detector spectra in these two programs}
\label{pbeam_gnumi_spectra}
\end{figure}

We have used these two  programs  corresponding to somewhat different 
beamline configurations: geometry of the target cave, details of the
magnetic field in the horn, thick window of the decay volume are among
the known differences between them. It is not surprising, therefore,
that the shape of the neutrino spectrum resulting from these two programs
are not identical, as shown in Fig.\ref{pbeam_gnumi_spectra}. In a similar
fashion these two programs yield different spectra for the far detector.

From the spectrum at the near detector the far detector spectrum can be
calculated using the Eq.\ref{far_prediction}. A comparison of this prediction 
with the actual far detector, PBEAM-calculated, spectrum is shown in Fig.
\ref{pbeam_reconstructed} together with the similar comparison for the 
'double ratio' method.  Despite the major differences in the simulation
programs and the resulting difference in the predicted neutrino spectra
between the program used to derive the correlation matrix and the program
used to 'analyze' the data the prediction is good to a level $\sim5\%$
or better. Again, the 'double ratio' is more sensitive to the differences
between the program than the matrix method.

\afterpage{\clearpage
\begin{figure}[H]
\centerline{\epsfig{file=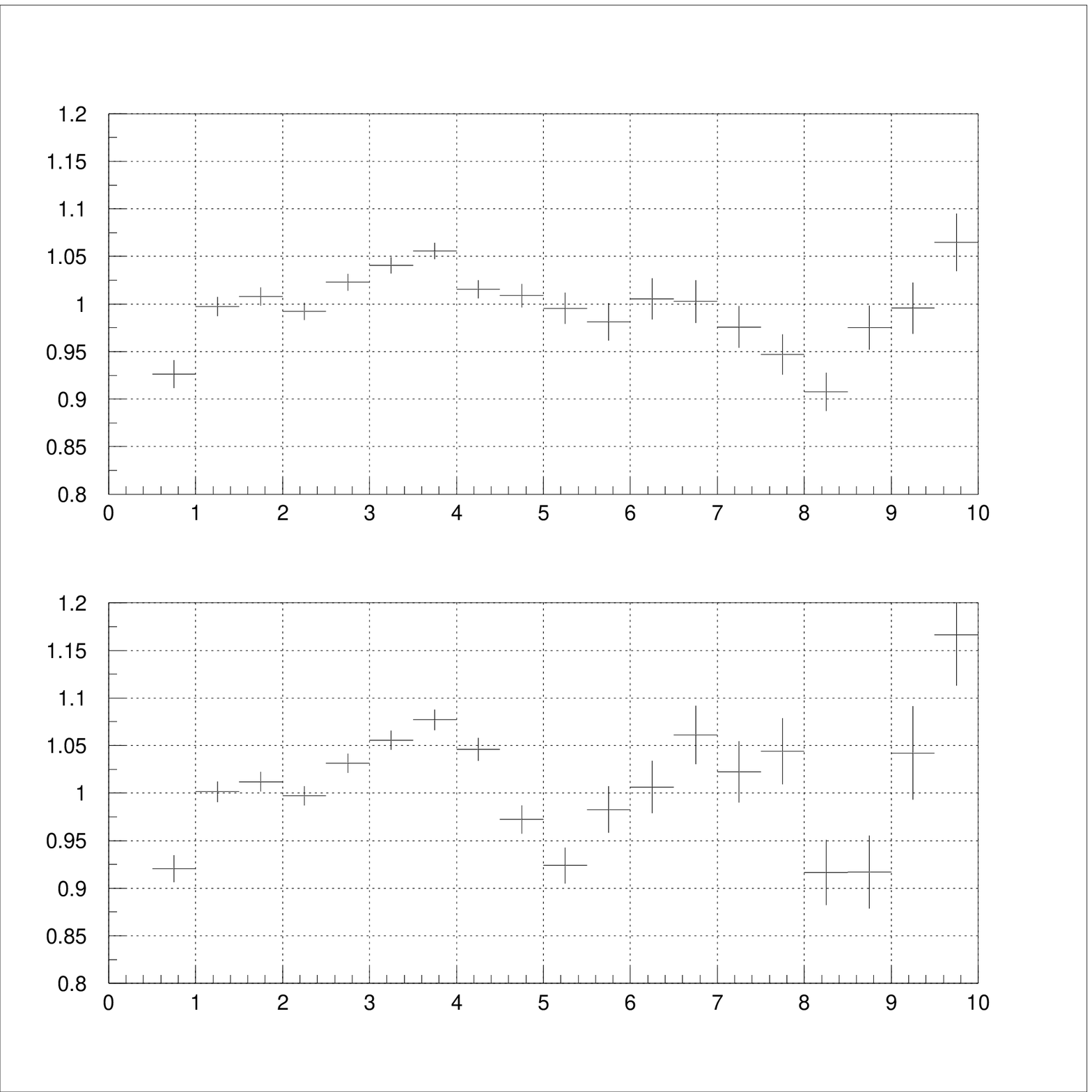,height=11cm}}
\caption{Up: ratio of the  far detector spectrum derived using GNUMI-based
correlation matrix to the actual one as predicted by PBEAM.
Down: the same  ratio for the far detector spectrum predicted using 
a 'double ratio' method.}
\label{pbeam_reconstructed}
\end{figure}
}
\subsection{Distribution of Pions  in the Decay Volume}

As shown in previous sections, the difference between the near and far
detector spectra is directly related to the finite decay pipe size and
to the distribution of  decay vertices  inside the decay volume.

A robustness of the prediction of the far spectrum can be investigated
by artificially enforcing major deviations from the spatial distributions
produced by the beam simulation programs. While such a procedure
does not correspond, probably, to any realistic imperfections of the
beam line, it nevertheless offers an insight into a sensitivity of the method.

A radial beam profile can be modified by  applying a weight factor of the
form
\begin{equation}
w_{r}(r) = 1 + a \cdot [ \sqrt{x^2+y^2} - r_0 ] / r_0
\label{r_weight}
\end{equation}

\noindent
where $r_0 = 50$ cm corresponds to a  half-radius of the decay pipe  
and $a = 0.2, 0.5, -0.2, -0.5$ is the magnitude of the distortion.

\begin{figure}[h]
\centerline{\epsfig{file=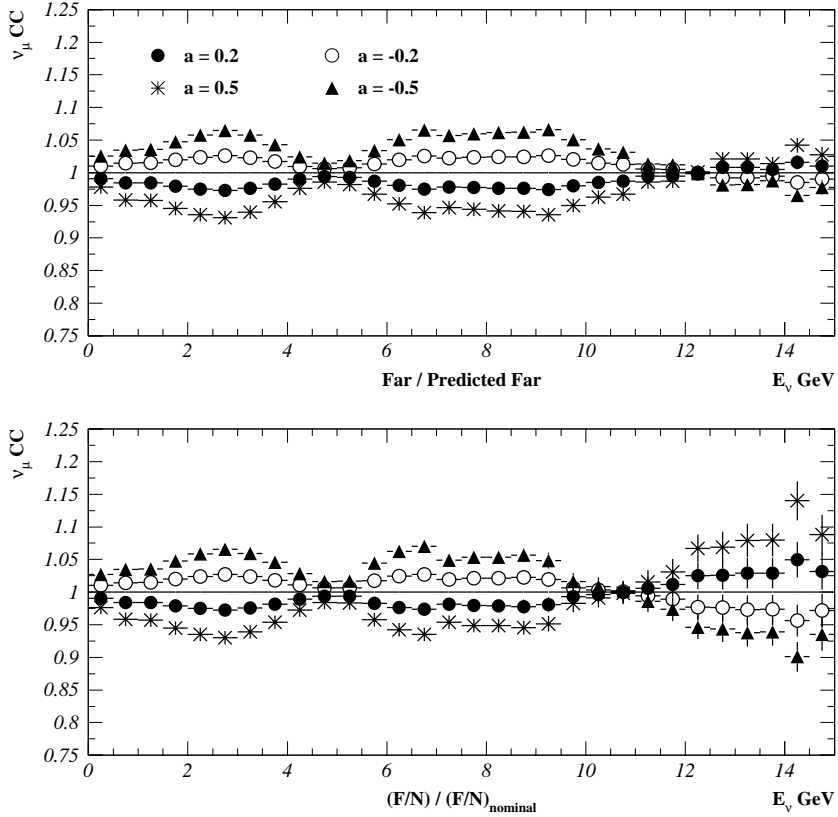,width=11cm}}
\caption{Up: ratios of far spectra for different beam profiles
to those predicted from the corresponding near spectra using
the same ``nominal" $M$ matrix.  Beam profiles were defined by changing
parameter $a$ in formula (\ref{r_weight}).
Down: an equivalent
comparison of Far/Near ratios over the nominal Far/Near ratio.}
\label{radial_distortion}
\end{figure}

Application of the standard transformation 
to the near detector neutrino spectrum corresponding to the shrunk down or blown-up
hadron beam leads to a prediction of the far detector spectrum with the accuracy
 shown in Fig.\ref{radial_distortion}.

Prediction of the shape of the neutrino spectrum in the far detector is not very
sensitive to the details of the transverse beam profile.
Shrinking and expanding the transverse profile of the hadron beam by
$\pm20\%$ affects the predicted far detector spectrum by $2\%$ or less.
Even more drastic change of the profile by $\pm50\%$ leads to a modest
$\leq5\%$ change in the predicted spectrum.

Variation of the distribution of the decay vertices along the beam axis is potentially 
more serious source of the systematic error, given the proximity of the near detector.
A sensitivity to the knowledge of this distribution can be evaluated
in a similar manner by re-weighting the decay points with
\begin{equation}
w_{z}(z) = 1 + a \cdot [ z - z_0 ] / z_0
\label{z_weight}
\end{equation}
\noindent
where $z_0 = 300$ m and $a$ is variable parameter, as before.
\begin{figure}[h]
\centerline{\epsfig{file=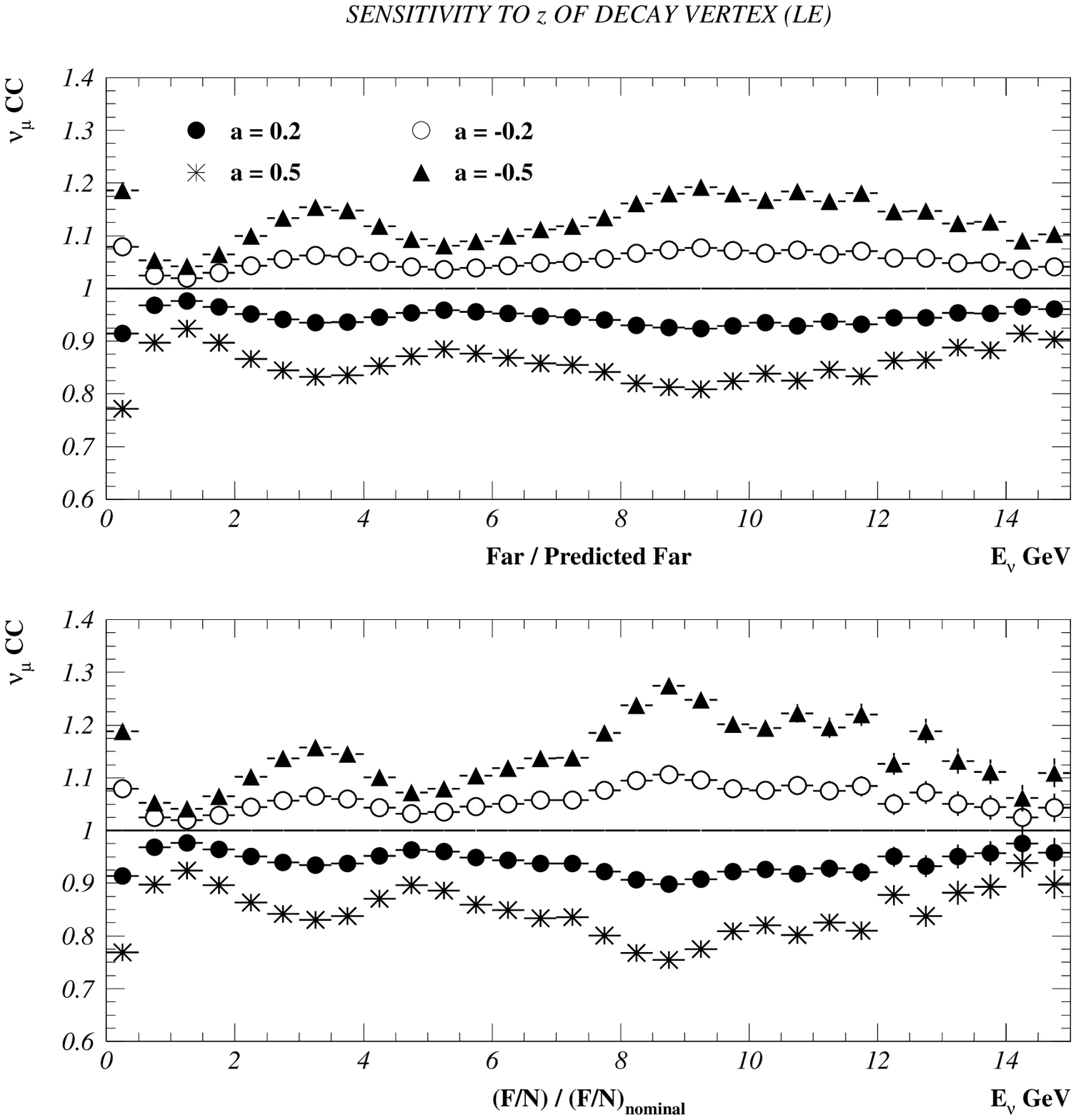,width=11cm}}
\caption{Up: ratios of far spectra for different pion decay vertex
distributions
to those predicted from the corresponding near spectra using
the same ``nominal" $M$ matrix.  Vertex distributions were modified by
changing
parameter $a$ in formula (\ref{z_weight}).
Down: an equivalent
comparison of Far/Near ratios over the nominal Far/Near ratio.}
\label{z_distortion}
\end{figure}

Results, shown in Fig.\ref{z_distortion}, indicate that the knowledge of the 
longitudinal distribution of the decay points to $\pm20\%$ is needed to keep
the systematic error within the bounds of the statistical accuracy. It is not
clear what effects in the beam line may cause such a significant re-shuffling
of the decay points. One possible source of such an effect would be an 
additional component of the hadron beam consisting of well collimated pions
thus traveling much longer distance along the decay volume. Such a component
can be created, for example, by a proton beam scraping inside of the collimators
upstream of the target. High energy pions created in such interactions
and focused by magnetic horn would have small angular divergence (owing to
the large longitudinal momentum component) and hence they would produce
disproportionally large neutrino flux in the near detector.    

To keep the systematic errors under control it is therefore very important
to minimize the amount of materials which can be intercepted by the proton beam
upstream of the target. It is perhaps possible to constrain the actual distribution
of the hadron in the decay pipe , and thus reduce the systematic error,by monitoring 
the energy deposition along the length of the decay volume. 

\subsection{Horn Current}  

The event rates and their energy distribution in the far and near detectors, 
as well as the far/near ratio depends on the strength of the focusing elements.
These effect was investigated using samples of events generated with GNuMI
the horns current changed from the nominal value
of $200\,kA$ to the values of $180\,kA$ and $220\,kA$.
The near detector spectra were used to compute the far detector spectra using
the matrix $M$ corresponding to the nominal horn current. 

\begin{figure}[h]
\hspace{1.5cm}\epsfig{file=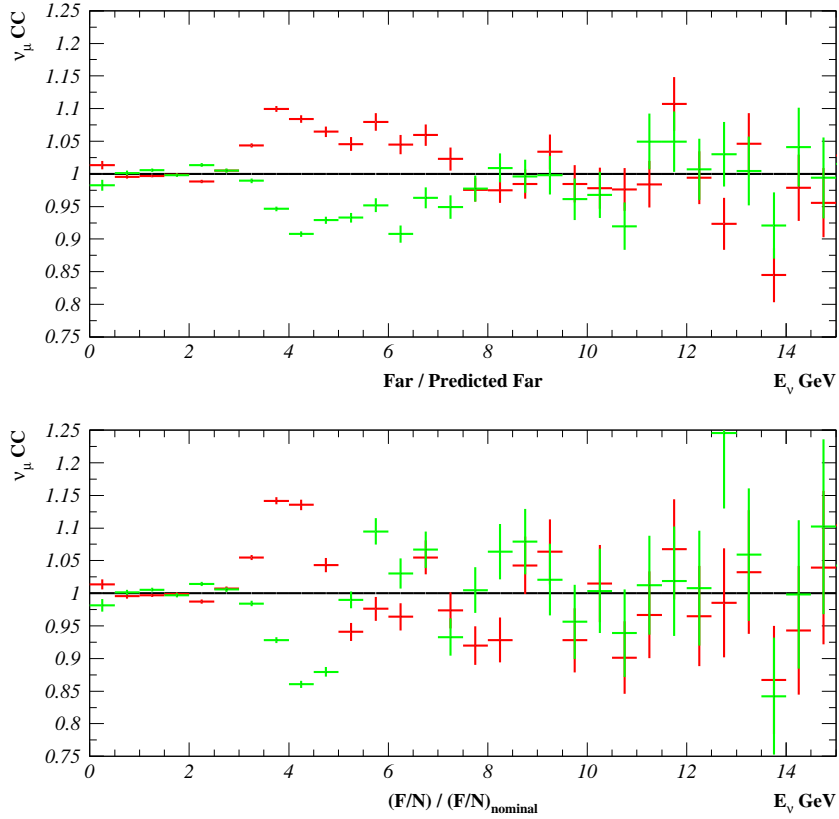,width=11cm}
\caption{Up: ratios of far spectra for different horn currents
(red: I = 0.18 MA, green: I = 0.22 MA) and those predicted from the
corresponding near spectra using the same ``nominal" $M$ matrix.
Down: an equivalent
comparison of Far/Near ratios over the nominal Far/Near ratio.}
\label{horn_current}
\end{figure}

Fig.\ref{horn_current} shows that to maintain  the neutrino flux prediction
with the accuracy below $5\,\%$ it is necessary to monitor the horn current
to $5\%$ as well. In addition to the monitoring of the peak current
for the horn, the timing of the current pulse must be monitored as well.
In a particular case of the NuMI horn with $1.7\,msec$ pulse it is required
that the timing of the pulse is monitored to better than $150\,\mu sec$ to maintain
the required level of the systematic error.

\subsection{Horn Displacements}

\begin{figure}[h]
\hspace{1.5cm}\epsfig{file=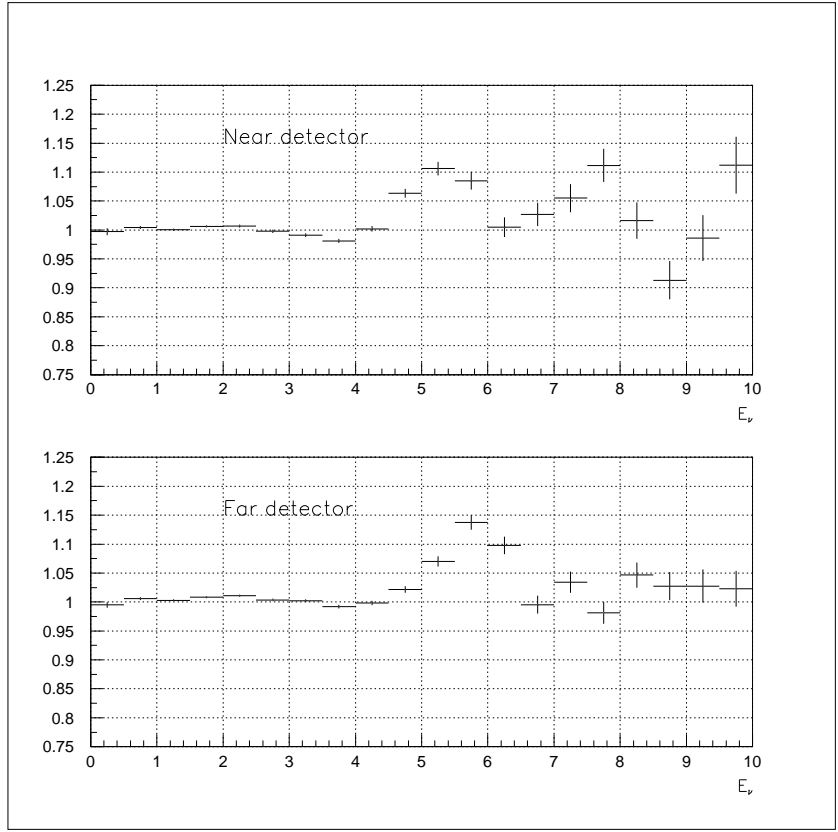,width=11cm}
\caption{Ratios of the event spectra in the near (top) and far (bottom) 
detectors observed with the horn 1 displaced by $2\,mm$ to the spectra
with horn 1 in the nominal position.}
\label{horn_displaced_spectra}
\end{figure}

Position of the magnetic horns defines edges of the beam acceptance and
the position (in the phase space of the produced particles) of the boundary 
between the focused beam and the bare target beam. Small displacement of the
horn causes some bin of high momentum  pions to be focused and increase
significantly their contribution to the neutrino flux, at the expense
of the small fraction of neutrinos at slightly lower energy. Displacement
of the first horn will produce also change of the phase space of particles
focused by the second horn, too. The resulting change of the event spectra
observed in the near and far detectors is shown in Fig.\ref{horn_displaced_spectra}.

Due to the finite radial extent of the decay volume the additional component
of the neutrino flux shows up at slightly different energies in the near and far
detectors. As a result, the sensitivity of the 'double ratio' method to
the horn displacement is somewhat enhanced.

\begin{figure}[h]
\hspace{1.5cm}\epsfig{file=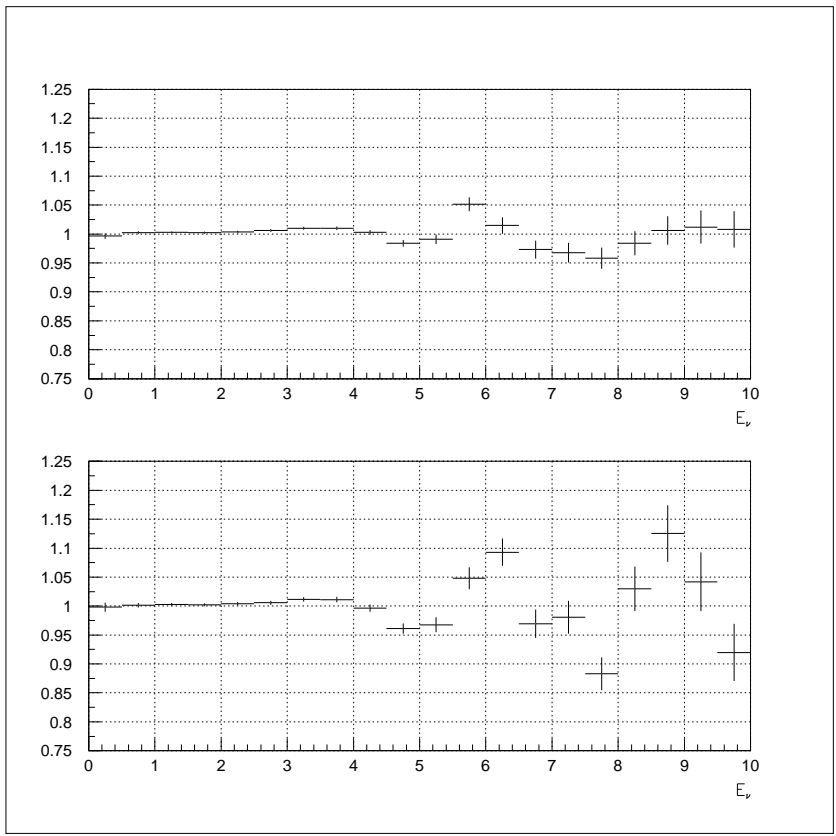,width=11cm}
\caption{Ratios of the predicted and observed spectra in the far detector 
when horn 1 is displaced laterally by 2 mm. Top figure is for the prediction 
using the matrix $M$ corresponding to the nominal horn position, bottom figure
is for the double ratio' method.}
\label{horn_displaced_prediction}
\end{figure}

The far detector spectrum predicted with the help of the correlation matrix $M$,
Eq.\ref{far_prediction} is much less sensitive to the horn displacements, as 
illustrated in Fig.\ref{horn_displaced_prediction}.

\section{Further Improvements: Large Near Detector}

The correlation matrix $M$ defined in Eq.\ref{matrix} relates the near and far detector
spectra averaged over the detectors volume. The far detector, in all practical cases,
is located so far from the neutrino source that there is no variation of the neutrino
spectrum over the area of the detector.

The transverse size of the MINOS near detector is small compared to the typical
beam size of the low energy beam, hence the near detector spectrum is practically identical
over the entire fiducial volume.

In case of a near detector large enough to detect spatial variation  of the neutrino
flux the correlation matrix can be further refined to take advantage of the additional
information. This might be the case, for example, for the K2K experiment.

A correlation matrix $M_{dA}$ can be calculated by carrying out the integral \ref{matrix}
using the weighting function ${W_{dA}(z,r_{dec},\theta,p,z_{n},E_{n})}$ describing
the  probability that
a pion(kaon) with momentum $p$ and the angle $\theta$ decaying at the position $z$ 
along the decay volume at the radius $r_{dec}$ will produce neutrino with
energy $E_{near}$ inside the detector area $dA$ in the near detector.

The far detector spectrum can be computed by integration over the near detector
area of the predicted contributions, as implied by the spectrum of events detected in
the area elements $dA$:
 
\begin{equation}
\left( F_{1},F_{2},\cdots,F_{n} \right) = 
\int dA
\begin{bmatrix}
M_{11}&M_{12}&\dots&M_{1n}\\
M_{21}&M_{22}&\dots&M_{2n}\\
\hdotsfor{4}\\
M_{n1}&M_{n2}&\dots&M_{nn}\\
\end{bmatrix}_{dA}
\begin{bmatrix}
N_{1}\\
N_{2}\\
\hdotsfor{1}\\
N_{n}
\end{bmatrix}_{dA}
\label{far_prediction_general}
\end{equation}

where the vector $N_{dA} = \left( N_{1},N_{2},\cdots,N_{n} \right)_{dA}$ is the spectrum
of the events detected in the area $dA$ of the near detector.

\section{Conclusions}

A prediction of the neutrino flux in the far detector from the one observed
in the near detector is very robust. This prediction is primarily determined
by the geometry of the beam line and the properties of the focusing elements.

The overall event rate can be predicted
with an accuracy better than $2\%$. Systematic error on the shape of the energy 
distribution of neutrinos are of the order of $2-3\%$ for the  main part of the 
flux (corresponding to the beam created by the well focused fraction of the
hadron beam) and  of the order $5-10\%$ in the tails.

The sensitivity of the far detector flux prediction is maximal in the energy
regions corresponding to the edges of the acceptance of the magnetic horns.
In a particular case of the low energy NuMI beam it is a region around $5\,GeV$
and $8-10\,GeV$, corresponding to the acceptance edges of the horn 1 and horn 2,
respectively.

An improved prediction method, utilizing the correlation of the energy spectra,
reduces the sensitivity of the far detector spectrum prediction, especially
to the effects producing localized distortions of the spectrum. This method enables
a reliable prediction of a neutrino flux at different distant locations, based
on the same near detector measurement.  

\section{Acknowledgments}

Numerous discussions with our collegues from NuMI/MINOS have helped us to understand
the issues presented in this paper. We are particularily grateful to A. Byon, J. Hylen
and A. Marchionni for their critical review of this paper.
 
 This work was supported in part by grants from the Illinois Board of
 Higher Education, the Illinois Department of Commerce and Community
 Affairs, the National Science Foundation, and the U.S. Department of
 Energy.


\newpage
\begin{thebibliography}{3}
\vspace*{1cm}
\bibitem{CCFR} I.E. Stockdale et al., Phys. Rev. Lett. 52, 1384 (1984)
\bibitem{CDHSW} F. Dydak et al., Phys. Lett. B134, 281, (1984)
\bibitem{CHARM} F. Bergsma at. al., Phys. Lett. B142, 103, (1984)
\bibitem{K2K} K2K Collaboration, to be published in Phys. Lett. B(2001);

K. Nishikawa et al., KEK-PS proposal (E362) (1995)
\bibitem{Atherton} H.W. Atherton et al., CERN 80-07, 1980
\bibitem{SPY1} G. Ambrosini et al., Phys. Lett. B420, 225, (1998);

G. Ambrosini et al., Phys. Lett. B425, 208, (1998)
\bibitem{SPY2} G. Ambrosini et al., Eur. Phys. J. C10, 605, (1999) 
\bibitem{Barton} D.S. Barton et al., Phys. Rev. D35, 35, (1987)
\bibitem{FLUKA} P.A. Aarnio et al., CERN TIS-RP-190 (1987)
\bibitem{MARS} N.V. Mokhov and S.I. Striganov, AIP Coinf. Proc. 435, 543 (1997)
\bibitem{Malensek} A.J. Malensek, Fermilab Report FN-341 (1981)
\bibitem{BMPT} M. Bonesini, A. Marchionni, F. Pietropaolo, T. Tabarelli de Fatis,
Eur. Phys. J. C20, 13-27, (2001)
\bibitem{Alberto} Alberto Marchionni, private communication.
\bibitem{hadron} M. Messier et al., {\it ``Neutrino Fluxes, Hadron Production,
and the Hadronic Hose"} (NuMI-B-700);
\bibitem{myhadr} M. Szleper, {\it ``Study of Hadron Production Models with
Hadron Monitors"} (NuMI-B-762).
\bibitem{gokhan} The Ntuples were generated at Nortwestern University by
Gokhan Unel. They correspond to a version of GNUMI as of spring 2001.
\bibitem{MMessier} M. Messier, Harvard University, private information
\addcontentsline{toc}{section}{References}



\end{thebibliography}
\end{document}